\documentclass[10pt, aps, prd, amsmath, floats, floatfix, twocolumn, notitlepage,
superscriptaddress, nofootinbib, showpacs, longbibliography]{revtex4-2}
\usepackage{aas_macros}
\usepackage{tikz}
\usepackage{pgfplots}
\pgfplotsset{compat=1.15}
\usepackage[utf8]{inputenc}
\usepackage[T1]{fontenc}
\usepackage{comment}
\usepackage{bm}
\usepackage[normalem]{ulem}\usepackage{mathtools,amsmath,amssymb,amsfonts,mathrsfs,eucal,graphicx,tensor,csquotes,accents,commath,chngcntr,siunitx}
\usepackage[dvipsnames]{xcolor}
\usepackage[unicode]{hyperref}
\usepackage{orcidlink}
\hypersetup{colorlinks=true, citecolor=MidnightBlue,
            linkcolor=MidnightBlue, urlcolor=MidnightBlue, linktocpage=true}
\usepackage[normalem]{ulem}

\DeclareMathAlphabet{\mathpzc}{OT1}{pzc}{m}{it}

\definecolor{darkgreen}{rgb}{0.0, 0.6, 0.0}

\newcommand{\JHU}{William H. Miller III Department of Physics and Astronomy, Johns Hopkins University, 3400 North Charles Street, Baltimore, Maryland, 21218, USA}

\newcommand{\IUCAA}{Inter-University Centre for Astronomy and Astrophysics, Post Bag 4,  Pune 411007, India}

\newcommand{\IIA}{Indian Institute of Astrophysics, Block 2, 100 Feet Road, Koramangala, Bengaluru 560034, India}

\newcommand{\IA}{Instituto de Astrofísica e Ciências do Espaço, Faculdade de Ciências da Universidade de Lisboa, Edifício C8, Campo Grande, P-1749-016 Lisbon, Portugal}

\newcommand{\LU}{Departamento de Física, Faculdade de Ciências da Universidade de Lisboa, Edifício C8, Campo Grande, P-1749-016 Lisbon, Portugal}

\labelformat{section}{Section #1} 
\labelformat{subsection}{Section #1} 
\labelformat{subsubsection}{Section #1}
\labelformat{subsubsubsection}{Section #1}
\labelformat{equation}{Eq.~(#1)} 
\labelformat{figure}{Fig.~#1} 
\labelformat{subfigure}{Fig.~\thefigure#1} 
\labelformat{table}{Table~#1} 
\labelformat{appendix}{Appendix #1}
\begin{document}

\date{\today}

\title{From Density to Mass: A New Parametrized Framework for Dark Matter Environment Around Black Holes}
\author{Rajes Ghosh~\orcidlink{0000-0002-1264-938X}}
\email[]{rghosh13@jh.edu}
\affiliation{\JHU}
\author{Nicholas Speeney~\orcidlink{0000-0002-2738-0985}}
\email[]{nspeene1@jhu.edu}
\affiliation{\JHU}
\author{Avijit Chowdhury~\orcidlink{0000-0002-7235-5076}}
\email[]{avijit.chowdhury@iiap.res.in}
\affiliation{\IIA}
\author{Chiranjeeb Singha~\orcidlink{0000-0003-0441-318X}}
\email[]{chiranjeeb.singha@iucaa.in}
\affiliation{\IUCAA}

\author{Miguel A. S. Pinto~\orcidlink{0000-0002-1327-0996}}
\email[]{mapinto@ciencias.ulisboa.pt}
\affiliation{\JHU}
\affiliation{\IA}
\affiliation{\LU}
\author{Emanuele Berti\orcidlink{0000-0003-0751-5130}}
\email[]{berti@jhu.edu}
\affiliation{\JHU}

\begin{abstract}
Astrophysical black holes (BHs) are not isolated vacuum objects but are expected to reside within complex environments, such as dark matter (DM) halos and baryonic distributions. Such environments can modify the spacetime geometry surrounding the BH and leave observable imprints on gravitational-wave (GW) signals, including shifts in the quasinormal mode (QNM) spectrum, tidal Love numbers (TLNs), and the inspiral dynamics of extreme-mass-ratio inspirals (EMRIs). We develop a general framework for describing DM-dressed BH spacetimes by parameterizing the geometry directly in terms of the enclosed mass function, rather than the traditional density-based profile. This mass-based formulation is naturally connected to the total gravitational field, accommodates several common halo profiles as limiting cases, and incorporates basic physical consistency requirements such as regularity, causality, and appropriate asymptotic behavior. Within this framework, we investigate the environmental imprints on QNMs, TLNs, and GW fluxes from EMRIs. We find a complementary sensitivity to the halo structure: QNMs predominantly probe the inner regions of the environment, whereas static TLNs are more sensitive to its outer structure. EMRI fluxes, in contrast, probe the matter distribution through its influence on the orbital motion and GW propagation. Our results establish the enclosed-mass formulation as a flexible and physically controlled framework for connecting astrophysical environment models with precision GW observables.
\end{abstract}

\maketitle
\section{Introduction}
In contrast to their idealized vacuum counterparts, astrophysical black holes (BHs) are seldom isolated and instead inhabit complex environments, ranging from stellar-mass BHs embedded in dense star clusters and gas-rich disks to supermassive BHs at the centers of galaxies~\cite{j_binney_galactic_1987, 1996Morris}. In particular, substantial observational and theoretical evidence indicates that dark matter (DM) halos surround galactic nuclei, potentially clustering around BHs and extending well beyond the visible extent of host galaxies~\cite{j_binney_galactic_1987, 1996Morris, mo2010galaxy}. These environments are further enriched by baryonic components such as stars, gas, compact remnants, and may also include multiple DM species with distinct clustering properties~\cite{Feng:2010gw, Alexander:2016aln, Ferreira:2020fam}. Realistic BHs should therefore be viewed not as vacuum solutions of general relativity (GR), but as objects embedded in extended astrophysical environments that constitute their natural habitat.

The presence of DM halos may significantly influence the formation and subsequent dynamics of compact binaries, thereby providing a natural laboratory to probe the yet-elusive properties of DM~\cite{Barausse:2014tra,  Cardoso:2021wlq, Speeney:2022ryg,Zhao:2023tyo,Speeney:2024mas,Chakraborty:2024gcr,DOnofrio:2026ulh, Chowdhury:2025tpt, Pezzella:2024tkf, Kavanagh:2020cfn,Duque:2023seg,Barausse:2014pra,Zhang:2024ugv,Gliorio:2025cbh,Mitra:2025tag,Fonseca:2025ehf, Macedo:2024qky,Bhowmik:2026owi}. Such environmental effects can leave cumulative imprints on both the generation and propagation of gravitational waves (GWs), most notably through inspiral dephasing induced by dynamical friction and gradual modifications of the effective binding potential~\cite{Barausse:2014pra,Barausse:2014tra,Cardoso:2019rvt}. In addition, the surrounding matter distribution modifies the effective spacetime of the remnant BH, leading to shifts in the characteristic quasinormal mode (QNM) spectrum relative to the isolated case~\cite{Barausse:2014tra, Cardoso:2021wlq, Speeney:2024mas, Chakraborty:2024gcr,Pezzella:2024tkf,Bhowmik:2026owi}. With the rapid advancement of GW astronomy, these effects may become accessible through increasingly precise waveform modeling~\cite{Amaro-Seoane:2012lgq, Babak:2017tow}. The prospect is particularly promising for extreme-mass-ratio inspirals (EMRIs), where the large number of in-band orbital cycles enables exceptional sensitivity to environmental perturbations~\cite{Barausse:2014tra, Kavanagh:2020cfn,Duque:2023seg,Barausse:2014pra,Zhang:2024ugv,Gliorio:2025cbh,Mitra:2025tag}.

Traditionally, the gravitational influence of DM in astrophysical systems is modeled through a prescribed density profile $\rho(r)$, motivated by cosmological $N$-body simulations and observational constraints~\cite{Navarro:1996gj, 1969Afz.....5..137E, Hernquist:1990be, Burkert:1995yz}. On galactic scales, a range of profiles provides accurate descriptions of different environments. The Navarro–Frenk–White (NFW) profile captures the universal behavior of collisionless DM halos seen in cosmological simulations~\cite{Navarro:1996gj}, while phenomenological models like the Einasto profile are used to describe galactic DM halos~\cite{1969Afz.....5..137E}. In stellar bulges and elliptical galaxies, Hernquist-type profiles offer an analytic description of observed mass and light distributions~\cite{Hernquist:1990be}. In dwarf and low-surface-brightness galaxies, observations of rotation curves and stellar kinematics can favor cored configurations, empirically modeled by the Burkert profile~\cite{Burkert:1995yz}. However, the adiabatic growth of a central BH (especially intermediate-mass and supermassive BHs relevant to galactic nuclei) within a surrounding DM distribution can lead, according to both Newtonian and relativistic analyses, to the formation of a steep DM cusp or spike. This enhancement comes with a characteristic cutoff scale, $r_0$, below which the density is strongly suppressed and tends to vanish near the horizon~\cite{Gondolo:1999ef, Sadeghian:2013laa, Ferrer:2017xwm, Speeney:2022ryg, Speeney:2024mas}.

While this density-based framework is physically well-motivated and successful, its inference from observables is typically indirect. Astrophysical measurements, such as stellar kinematics, gas rotation curves, and gravitational lensing, primarily probe the cumulative gravitational field or, equivalently, the enclosed mass function $m(r)$, rather than the full functional form of the density profile $\rho(r)$ itself. Although specifying $\rho(r)$ and $m(r)$ is mathematically equivalent via Einstein’s equations, phenomenologically modeling $m(r)$ is both more natural and directly connected to observations. In addition, an $m(r)$-based formulation naturally accommodates multi-component environments, including cold and ultralight DM as well as baryonic matter, by simply summing the contributions of the individual non-interacting components, whose only coupling is gravitational, without requiring any assumptions about their species-specific microphysics. From an observational standpoint, this is particularly relevant because standard astrophysical probes are largely insensitive to how the mass is partitioned among different subpopulations~\cite{j_binney_galactic_1987,Poisson:2009pwt,Persic:1995ru,Navarro:1996gj,Bertone:2016nfn,2010MNRAS.406.1220W,1992grle.book.....S, 1969Afz.....5..137E}. 

Motivated by these considerations and other reasons discussed in subsequent sections, we aim to develop a complementary framework in which the gravitational field is directly parametrized in terms of the enclosed mass profile $m(r)$. By construction, this formulation encodes the coarse-grained properties most directly probed by observations, while providing a flexible and unified description of the matter distribution. Standard density-based approaches are often formulated in terms of specific functional forms, such as power-law profiles like NFW or exponential profiles like Einasto, each tailored to particular physical scenarios. As a result, analyses frequently proceed on a case-by-case basis, with limited ability to systematically interpolate between different models. The enclosed mass formulation, however, naturally unifies these descriptions, with features of various density profiles emerging as limiting cases. Moreover, our $m(r)$-based approach directly incorporates physical consistency considerations, like regularity, energy conditions, and causality, thereby offering a controlled framework for modeling DM-induced BH spacetimes.

An important motivation for this framework is the growing capability of GW observations to probe new physics with increasing precision. Environmental effects can leave their imprint on several observables, including the QNM spectrum of the remnant BH~\cite{Barausse:2014tra, Cardoso:2021wlq, Zhao:2023tyo, Speeney:2024mas, Chakraborty:2024gcr,Pezzella:2024tkf, Bhowmik:2026owi}, the tidal Love numbers (TLNs) characterizing its response to external perturbations~\cite{Chakraborty:2024gcr,  Cardoso:2019upw, Cardoso:2021wlq, Chakravarti:2025awj, DOnofrio:2026ulh, Zhao:2026eti,Chakraborty:2026qru,Cannizzaro:2024fpz,Chowdhury:2026cjv,Bhowmik:2026owi}, and GW fluxes for EMRIs~\cite{Barausse:2014tra, Kavanagh:2020cfn,Duque:2023seg,Barausse:2014pra,Zhang:2024ugv,Gliorio:2025cbh,Mitra:2025tag}. In particular, deviations from vacuum GR can be directly mapped to features of the underlying mass profile modeled by $m(r)$. We provide a systematic analysis of these effects and demonstrate how our formulation offers a conceptually clean and observationally grounded route to incorporating environmental physics in BH spacetimes. While the present work focuses on spherically symmetric configurations, the framework can be extended to rotating systems in a straightforward manner. Throughout the paper we use geometrical units ($G=c=1$).

\begin{figure}[h!]
\centering
\begin{tikzpicture}

\begin{axis}[
    axis x line=bottom,
    axis y line=none,
    xmin=0, xmax=11,
    ymin=0, ymax=1,
    xtick={1.2,3.6,6.1,8.5},
    xticklabels={$r_H$,$r_0$,$r_s$,$r_c$},
    ytick=\empty,
    width=1.17\linewidth,
    height=5cm,
    enlargelimits=false,
    clip=false,
    axis line style={line width=1.5pt,-},
]


\addplot [draw=none, fill=black] coordinates {(0,0) (1.2,0) (1.2,1) (0,1)};
\node[text=white] at (axis cs:0.6,0.5) {\small BH};

\addplot [draw=none, fill=gray!25] coordinates {(1.2,0) (3.6,0) (3.6,1) (1.2,1)};
\node[align=center] at (axis cs:2.4,0.5) {\small Near-horizon\\Region \\ \footnotesize(Vacuum)};

\addplot [draw=none, fill=blue!20] coordinates {(3.6,0) (6.1,0) (6.1,1) (3.6,1)};
\node[align=center] at (axis cs:4.85,0.5) {\small Inner\\ Region\\ \footnotesize($r_0 \ll r \ll r_s$)};

\addplot [draw=none, fill=green!20] coordinates {(6.1,0) (8.5,0) (8.5,1) (6.1,1)};
\node[align=center] at (axis cs:7.3,0.5) {\small Outer\\ Region\\ \footnotesize($r_s \ll r \leq r_c$)};

\addplot [draw=none, fill=gray!10] coordinates {(8.5,0) (11,0) (11,1) (8.5,1)};
\node[align=center] at (axis cs:9.75,0.5) {\small Asymptotic\\Region\\ \footnotesize(Vacuum)};

\draw[dashed] (axis cs:1.2,0) -- (axis cs:1.2,1);
\draw[dashed] (axis cs:3.6,0) -- (axis cs:3.6,1);
\draw[dashed] (axis cs:6.1,0) -- (axis cs:6.1,1);
\draw[dashed] (axis cs:8.5,0) -- (axis cs:8.5,1);

\node[above] at (axis cs:0.6,1) {\footnotesize $r < r_H$};
\node[above] at (axis cs:2.4,1) {\footnotesize $r_H \leq r \leq r_0$};
\node[above] at (axis cs:6,1) {\footnotesize DM Region: $r_0 \leq r \leq r_c$};
\node[above] at (axis cs:9.7,1) {\footnotesize $r > r_c$};

\node at (axis cs:10.6,-0.07) {$r \rightarrow$};

\end{axis}
\end{tikzpicture}

\caption{
Schematic radial structure (not to scale) of the DM distribution around a central BH.
The region $r \leq r_H$ corresponds to the BH interior. There is no DM in the
near-horizon region $r_H \leq r \leq r_0$. The DM distribution is characterized by an
\textbf{}inner (cusp) region $r_0 \ll r \ll r_s$, an outer (halo) region $r_s \ll r \leq r_c$, and an asymptotic vacuum region ($r > r_c$). As a limiting case, one may consider $r_c \to \infty$.
}
\label{fig:dm_regions}

\end{figure}
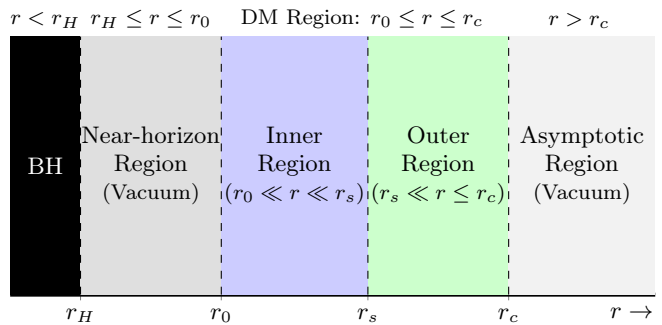

\section{Construction of the Model}
\label{sec:model_construction}
We consider a static, spherically symmetric, and asymptotically flat BH of mass $M_{BH}$ surrounded by a DM environment of total mass $M_E$. In vacuum GR, the BH spacetime is uniquely specified by the Schwarzschild metric everywhere due to Birkhoff's theorem~\cite{jebsen, birkhoff}. However, in the presence of the halo, the effective metric will take the form
\begin{equation} \label{metric}
    ds^2 =-f(r)\, dt^2+\frac{dr^2}{1-\frac{2\,m(r)}{r}}+r^2\, d\Omega^2_{(2)}~,
\end{equation}
where $d\Omega^2_{(2)}$ is the surface element of a $2$-sphere, $m(r) := (r/2)(1-g^{\mu \nu} \nabla_\mu r\, \nabla_\nu r)$ is the Misner–Sharp mass~\cite{Misner:1964je, Hayward:1994bu}, and $r_H=2M_{BH}$ is the location of the event horizon, where both $g_{tt}$ and $g^{rr}$ vanish~\cite{Vishveshwara1968}.

We consider the DM halo to be present within radii $[r_0,r_c]$, see \ref{fig:dm_regions}. Here, $r_0$ is the inner cutoff radius of the DM halo, which is usually set equal to the radius of the marginally bound orbit ($r_0=2r_H$) in the background Schwarzschild geometry following adiabatic arguments. However, we will keep it free and later constrain it using physical principles, like energy conditions and causality. On the other hand, $r_c \gg r_0$ is the outer cutoff radius of the DM halo, which we will also keep free and finite (but larger than any length scale of the system). In principle, $r_c$ can be set to be infinity, if allowed by the regularity of the halo. This can be achieved from our framework as a limiting case. 

Therefore, the domain of outer communication $[r_H,\infty)$ naturally breaks into three regions, namely the near-horizon region $[r_H,r_0]$, the DM region $[r_0,r_c]$, and the asymptotic region $[r_c,\infty)$. It is suggestive to write the metric components in \ref{metric} as
\begin{widetext}
\begin{equation} \label{metriccomp}
    \begin{aligned}
        &f(r) = f_s(r)\, e^{\mu_0}
        \left[1 - \Theta(r - r_0)\right]+ f_s(r)\, e^{\mu(r)}\left[\Theta(r - r_0) - \Theta(r - r_c)\right]+ \left(1 - \frac{2 M_{ADM}}{r}\right)
        \Theta(r - r_c)~,\\
        &m(r) = M_{BH}+m_E(r)
        \left[\Theta(r - r_0)- \Theta(r - r_c)\right] + M_E\,\Theta(r - r_c)~.
    \end{aligned} 
\end{equation}
\end{widetext}
Here, $f_s(r)=(1-r_H/r)$, $\Theta(x) = 0\, (1)$ for $x<0$ ($x \geq 0$) is the Heaviside unit-step function, and $M_{ADM}=M_{BH}+M_E$ is the Arnowitt–Deser–Misner (ADM) mass of the system. 

Note that for a realistic DM profile with no surface layers, a consistent matching in GR at a timelike hypersurface ($r=\text{const.}$) requires both the induced metric and extrinsic curvature to be continuous across different regions due to Darmois–Israel junction conditions~\cite{MSM_1927__25__1_0,Israel:1966rt,Poisson:2009pwt,Wald:1984rg}. For our case with \ref{metric} and \ref{metriccomp}, this is tantamount to matching $\{\mu(r)$, $\mu'(r)$, $m_E(r)\}$ at $r=\{r_0,r_c\}$, where $'$ denotes derivative with respect to the radial coordinate $r$. However, we will see in the next subsection that Einstein's equations algebraically connect $\mu'(r)$ with $m_E(r)$. Therefore, we only need to ensure the matching of $\mu(r)$ and $m_E(r)$ at $r=\{r_0,r_c\}$. Particularly at $r=r_0$, we must have 
\begin{equation} \label{matching}
    \mu(r_0)=\mu_0~,\quad m_E(r_0)=0~.
\end{equation}
The quantity $\mu_0$ can be fixed by matching $f(r)$ at $r_c$, i.e., by the condition $f_s(r_c)\, e^{\mu(r_c)}=(1-2M_{ADM}/r_c)$, and it quantifies the relative redshift factor between $r_H$ and $r_c$ caused by the DM halo. Hence, $\mu_0$ must be non-zero (in fact, it is negative) if we require $t$ to be the proper time at infinity. Indeed, the surface gravity at the horizon is given by $\kappa_H=e^{\mu_0/2}/(2r_H)$, which is a constant and obeys the zeroth law of BH mechanics~\cite{Bardeen:1973gs}. 

Moreover, since $r_c \gg r_0$, the DM density (mass) will become negligibly small (saturate to $M_E$) near the outer cutoff radius $r_c$. As a result, asymptotic matching requires $m_E(r_c)=M_E$. In order to smoothly incorporate the case with $r_c \to \infty$, we may consider
\begin{equation} \label{ADM}
    m_E(r \sim r_c \to \infty)\to M_E-\frac{c_1}{r^{n}}~,
\end{equation}
where $c_1$ is some positive constant (as $m_E(r)$ is a non-decreasing function) and some index $n > 0$. We have considered a power-law-type saturation, but one can, in principle, consider other functional forms as well.

\subsection{Implementing Einstein's equations and consistency criteria}
So far, we have only analyzed various kinematical properties of the metric given in \ref{metric}. Now, we will require it to be a solution of the non-vacuum Einstein's equations $G_{\mu\nu} = 8\pi\, T_{\mu \nu}^E$, where $T^E_{\mu \nu}$ denotes the environmental stress-energy tensor. Following the Einstein cluster prescription, we model $(T^E)^{\mu}_{\nu}=\text{diag.}[-\rho(r),0,p_t(r),p_t(r)]$ as an anisotropic fluid confined within $[r_0>r_H,r_c]$ with density $\rho(r)$, negligible radial pressure ($p_r \approx 0$), and tangential pressure $p_t(r)$. Then, the relevant Einstein's equations take the form
\begin{equation} \label{finalEin}
    \begin{split}
        &\rho(r) = \frac{m_E'(r)}{4\pi\, r^2}~,\, \, p_t(r)=\frac{1}{4} \left[r\, \mu'(r)+\frac{r_H}{r-r_H}\right] \rho(r)~,\\
        &\mu(r)=\mu_0+\int_{r_0}^r d\tilde r\, \frac{2\, m_E(\tilde r)}{(\tilde r-r_H)[\tilde r-r_H-2\, m_E(\tilde r)]}\,.
    \end{split}
\end{equation}
Since the DM profile starts at $r=r_0$, we will set $\rho(r_0)=0$. This, in turn, implies $m_E'(r_0)=0$ and $\mu'(r_0)=\mu''(r_0)=0$ by \ref{matching}. Also, following the discussion below \ref{matching} and assuming $r_c \gg r_0$ to be the largest DM scale, we can fix $\mu_0$ by the condition $\mu(r_c) \approx -2M_E/r_c$.

Now, we require that the DM profile must obey the dominant energy condition $|p_t(r)|/\rho(r) \leq 1$ with $\rho(r) \geq 0$~\cite{Alho:2021sli}, and causality condition $c_s^2(r) \leq 1/k$ (for some $k \geq 1$) on the sound-speed defined as $c_s^2(r) = \delta p_t/\delta \rho = p_t'(r)/\rho'(r)$~\cite{Datta:2023zmd,Zhao:2026eti,Pani:2025qxs}, especially in the limit $r \to r_0$. Both of these conditions imply
\begin{equation} \label{r0lower}
        r_0 \geq \left(\frac{k+4}{4}\right) r_{H}~. 
\end{equation}
Therefore, we are forced to set $r_0 > r_H$ for any $k$. In particular, the standard choice $r_0=2r_H$ is consistent with the above inequality for all $k \in [1,4]$. We should also keep in mind that \ref{r0lower} is only a necessary condition, but may not be sufficient (i.e., we may require an even stronger lower bound on $r_0$) as the dominant energy condition and causality should hold everywhere in $[r_0,r_c]$. Moreover, if we want to allow for a slow accretion of DM onto the BH, we must choose $r_0 < r_{ISCO} \approx 3r_H$ or $k<8$, where $r_{ISCO}$ is the radius of the innermost stable circular orbit of the system.

\begin{figure*}
\centering
\includegraphics[width=\linewidth]{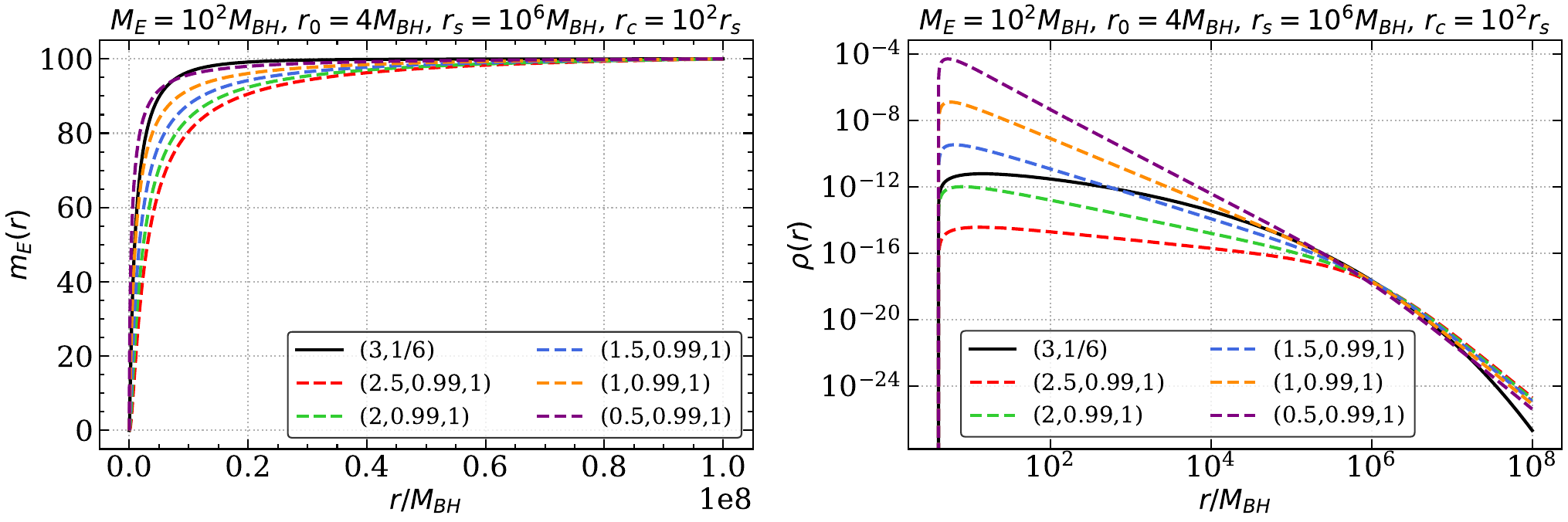}
\caption{Left: DM mass profile with radius $r \in [r_0=4M_{BH},r_c=10^8M_{BH}]$ for a fixed $\alpha=1$, $n_2=0.99$, $n_3=1$ and $n_1 \in \{0.5,1,1.5,2,2.5\}$. Note that the profile becomes steeper in the inner region with decreasing $n_1$, which is expected as $(r/r_s)\ll 1$ there. We have chosen the total DM mass $M_E=10^2M_{BH}$, transition scale $r_s=10^6M_{BH}$, and outer cutoff radius $r_c=10^2r_s$. The solid black line represents an illustrative case of \ref{1F1} with $n_1=3$, $n_3=1/6$ and $d=18$. Right: Corresponding density profile plots in log-log scale.}\label{fig2}
\end{figure*} 
\subsection{Building the model}
From \ref{finalEin}, it is evident that the system of equations effectively contains a single independent unknown. Traditionally, one chooses to specify the density profile $\rho(r)$ to close the system. However, this requires assuming a functional form of $\rho(r)$ over the entire range $[r_0,r_c]$, which is difficult to justify since astrophysical observations typically probe only the local slopes, rather than the full radial dependence. 

In contrast, observables at a given radius $r$ depend directly on the total enclosed mass, $m(r)=M_{BH}+m_E(r)$. For instance, in the Newtonian regime ($r \gg r_H$), the circular velocity of a star, planet, or gas cloud is $v^2(r) = m(r)/r$, while the gravitational potential is $\Phi_N(r) = -\int_{r}^{\infty} dr\, m(r)/r^2$. Moreover, as discussed in the Introduction, modeling $m_E(r)$ naturally provides a coarse-grained framework for scenarios where the environment is composed of multiple species, including DM and baryonic subpopulations. Motivated by these considerations, we adopt $m_E(r)$ as the primary quantity to model phenomenologically, guided by both physical arguments and observational inputs.

As discussed earlier, Einstein's equations along with matching conditions fix the behavior of $m_E(r)$ at $r_0$ as $m_E(r_0)=m_E'(r_0)=0$. Consequently, near the inner cutoff, the mass profile must scale as $m_E(r) \propto (1-r_0/r)^{1+\alpha}$ with $\alpha>0$. From observations, we also know that astrophysically relevant functions $m_E(r)$ involve a characteristic transition scale $r_s \gg r_0$. In the absence of such a scale, the DM mass would follow a simple power-law behavior ($\sim r^{p}$ with $p>0$) everywhere, reflecting scale invariance. However, realistic DM halos are shaped by astrophysical processes such as adiabatic growth and relaxation, which introduce a characteristic radius $r_s$ where the behavior of the system changes. Incorporating this scale via the dimensionless ratio ($r/r_s$) is therefore unavoidable to construct a physically meaningful profile. 

This naturally leads to two regimes, as shown in \ref{fig:dm_regions}: an inner region ($r_0 \ll r \ll r_s$) characterized by a scaling $(r/r_s)^{n_1}$ with $n_1>0$, and an outer region ($r \gg r_s$) of mass saturation. Accordingly, we model the DM mass profile as
\begin{equation} \label{mE1}
    m_E(r) = \frac{M_E}{N} \left(\frac{r}{r_s}\right)^{n_1}\left( 1-\frac{r_0}{r}\right)^{1+\alpha} \mathcal{T}\left(\frac{r}{r_s}\right)~,
\end{equation}
where $N$ is fixed by the normalization condition $m_E(r_c)=M_E$, and $\mathcal{T}(r/r_s)$ is a smooth transition function that interpolates between the inner and outer regimes. Demanding a smooth transition to the ADM condition at $r \to r_c$ (even when $r_c \to \infty$), \ref{ADM} can be rewritten as
\begin{equation*}
    m_E \sim \left(\frac{r}{r_s}\right)^{n_1} \left[M_E \left(\frac{r}{r_s}\right)^{-n_1}+c_2 \left(\frac{r}{r_s}\right)^{-(n_1+n)}\right]~,
\end{equation*}
where $c_2$ is a constant. The transition function $\mathcal{T}(r/r_s)$ must have two asymptotically decaying branches, as specified by the terms inside the square brackets. The most natural way to achieve this, while maintaining smoothness and regularity over the halo's parameter space, is to identify the transition function $\mathcal{T}(r/r_s)$ with Gauss's hypergeometric function ${}_2F_1(a,b,c;x)$~\cite{abramowitz1964handbook,NIST:DLMF,Arfken2012MathematicalMF,book:91229966} with suitable parameters. 

Thus, the complete parametrized model for $m_E(r)$ becomes
\begin{equation} \label{mE2}
    \begin{aligned}
        m_E(r) =\, &\frac{M_E}{N} \left(\frac{r}{r_s}\right)^{n_1}\left( 1-\frac{r_0}{r}\right)^{1+\alpha}\times \\ &\kern2em {}_2F_1\left(\frac{n_1}{n_3},\frac{n_1+n_2}{n_3},\frac{n_1+n_3}{n_3};-\left(\frac{r}{r_s}\right)^{n_3}\right)~,
    \end{aligned}
\end{equation}
where $N$ is fixed by demanding $m_E(r_c)=M_E$ and $\{\alpha,n_1,n_2,n_3\}>0$. This is the central equation in this work. For notational uniformity among the indices, we have relabeled the $n$-dependent index as $n_2$. The form of \ref{mE2} follows from the large-argument ($|x| \to \infty$) expansion of the Gauss hypergeometric function,  ${}_2F_1(a,b,c;x)\sim c_3\, (-x)^{-a}+c_4\, (-x)^{-b}$, for our parametrization. Finally, the argument $-(r/r_s)^{n_3}$ is chosen to be negative to avoid the singularity of Gauss's hypergeometric function at $r=r_s$. 

Some illustrative cases are plotted in \ref{fig2}.
\begin{figure*}
    \centering
    \includegraphics[width=0.49\textwidth]{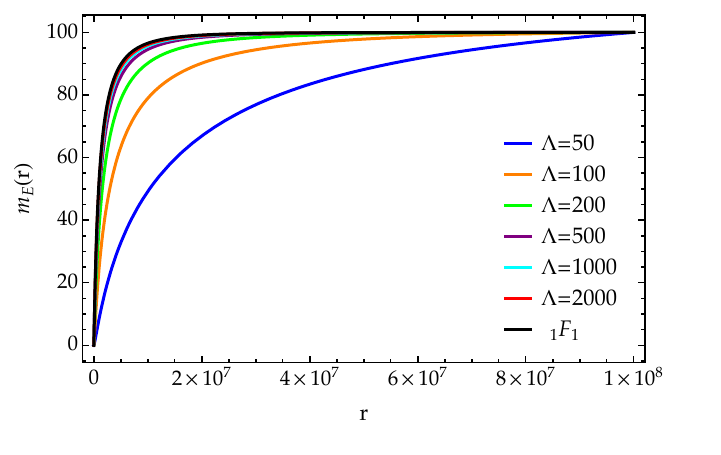}
    \includegraphics[width=0.49\textwidth]{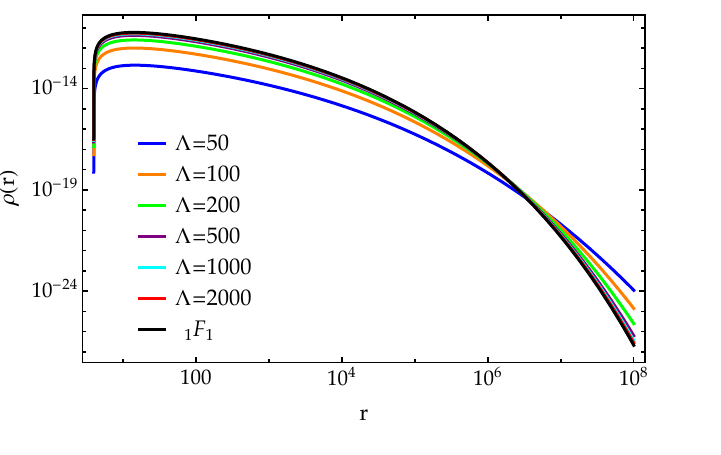}
    \caption{
    Left: Plot of $m_E(r)$ (in units of $M_{BH}=1$) with increasing $\Lambda = \{50,100,200,500,1000,2000\}$ in the ${}_2F_1$-based model as it approaches the limiting ${}_1F_1$-model. 
    Right: Corresponding density plot in log-log scale. We have fixed $n_1=3$, $n_3=1/6$, $d=18$, and all other parameters as mentioned in the caption of \ref{fig2}.}
    \label{fig2b}
\end{figure*}
A few comments are in order. Here, we adopt a power-law-type scaling (cusp) behavior in the inner region. Although the model readily generalizes to qualitatively different environments, such as cored DM profiles, we confine our attention to the cuspy scenario. Also, one can, in principle, consider more involved choices for the function $\mathcal{T}(r/r_s)$, but the above choice is both minimal and leads to an analytical handle on many derived quantities, as we will see below. 

In fact, although we have constructed the model by assuming a power-law saturation of $m_E(r)$ at large radii as specified by \ref{ADM}, the parameters in \ref{mE2} can be chosen judiciously to model other types of saturation as well. For instance, by setting $n_2 = (\Lambda\, n_3-n_1)$, and substituting $r_s \to r_s\, (\Lambda/d)^{1/n_3}$, we can reduce the ${}_2F_1$-function to a ${}_1F_1$-function for large values of $\Lambda$~\cite{abramowitz1964handbook,NIST:DLMF,Arfken2012MathematicalMF, book:91229966} using the relation:
\begin{equation} \label{1F1}
    \begin{aligned}
        &\lim_{\Lambda \to \infty} {}_2F_1\left(\frac{n_1}{n_3},\Lambda,\frac{n_1+n_3}{n_3};-\frac{d}{\Lambda}\left(\frac{r}{r_s}\right)^{n_3} \right) \\ 
        &\quad\quad = {}_1F_{1}\left(\frac{n_1}{n_3},\frac{n_1+n_3}{n_3};-d\left(\frac{r}{r_s}\right)^{n_3} \right)~.
    \end{aligned}
\end{equation}
Thus, by choosing larger and larger values of $\Lambda$ or equivalently of $n_2$, we can naturally produce an exponential saturation (combined with a power law, in general) of DM mass at large radii using the same profile as in \ref{mE2}. This is illustrated in \ref{fig2b} by plotting mass and density with increasing values of $\Lambda$ in \ref{mE2}. In other words, this ${}_1F_1$-model can be thought of as the limiting case of the more general ${}_2F_1$-model. This is a particularly useful feature of our model, providing sufficient mathematical flexibility to unify a range of seemingly distinct DM profiles.
\subsection{Parameter interpretations}
Various parameters of our mass model in \ref{mE2} possess interesting physical interpretations, which we will discuss in this section. To this end, we start by analytically computing the resultant density $\rho(r)$ from our mass model in \ref{mE2}. Using the first relation in \ref{finalEin}, we obtain 
\begin{widetext}
\begin{equation} \label{rho}
    \begin{split}
        \rho(r) = \frac{M_E}{4 \pi r_s^{3} N}\, \left(\frac{r}{r_s}\right)^{n_1-3} &\left(1-\frac{r_0}{r}\right)^{\alpha}\, \left[1+\left(\frac{r}{r_s}\right)^{n_3}\right]^{-(n_1+n_2)/n_3} \times \\
        &\Bigg[n_1 \left(1-\frac{r_0}{r}\right)+\frac{r_0(1+\alpha)}{r} \left\{1+\left(\frac{r}{r_s}\right)^{n_3}\right\} {}_2F_1\left(1,\frac{n_3-n_2}{n_3},\frac{n_1+n_3}{n_3};-\left(\frac{r}{r_s}\right)^{n_3}\right)\Bigg]~. 
    \end{split}    
\end{equation}
\end{widetext}
We note that the prefactor outside the square brackets closely resembles the so-called $(\alpha,\beta,\gamma)$-profile (this $\alpha$ should not be confused with the parameter $\alpha$ appearing in our profile)~\cite{Hernquist:1990be, Taylor:2002zd, 2020MNRAS.499.2912F}, with an inner cutoff at $r=r_0$. In fact, in the limit $r_0 \to 0$ with no BH at the center, this structural similarity with either an $(\alpha,\beta,\gamma)$-profile or an Einasto profile becomes more apparent, depending on the choice of $n_2$ being finite or tending to infinity as per the reparameterization discussed above \ref{1F1}. However, in general, the $r$-dependent factor in the square brackets cannot be absorbed inside other model parameters globally, distinguishing the resulting profile from the exact $(\alpha,\beta,\gamma)$/Einasto form. Besides, as discussed earlier, our mass-based formulation is more naturally connected to astrophysical observables and it offers a better analytical handle over various derived quantities.

Notably, there are other advantages of a mass-based DM model over density-based ones. To see this, let us derive a relation between the local density amplitude $\rho_0$ at $r_0$ and the net halo mass $M_E$ as
\begin{equation} \label{rho0}
    \rho_0 := \lim_{r\to r_0} (1-r_0/r)^{-\alpha} \rho(r) \approx \frac{(1+\alpha)\, M_E}{4\pi\, r_0^3\, N}\, \left(\frac{r_0}{r_s}\right)^{n_1}~,
\end{equation}
i.e., we have $\rho(r \to r_0) \to \rho_0 (1-r_0/r)^\alpha$. Here, we have used the fact that ${}_2F_1(a,b,c; x \to 0)=1$. Note that, in density-based models, specifying the halo mass $M_E$ requires a global normalization condition as in \ref{rho0}, or equivalently $4\pi \int_{r_0}^{r_c} r^2\, \rho(r)\, dr=M_E$. This implies that the local density amplitude $\rho_0$ must depend on the entire profile, mixing the fine-grained/coarse-grained properties of the halo. In contrast, the mass-based parametrization in \ref{mE2} is built using locally-defined quantities alone. This local construction also facilitates the incorporation of different regional features whenever needed, such as the presence of a cored DM structure.

\begin{figure}[h!]
\centering
\includegraphics[width=\linewidth]{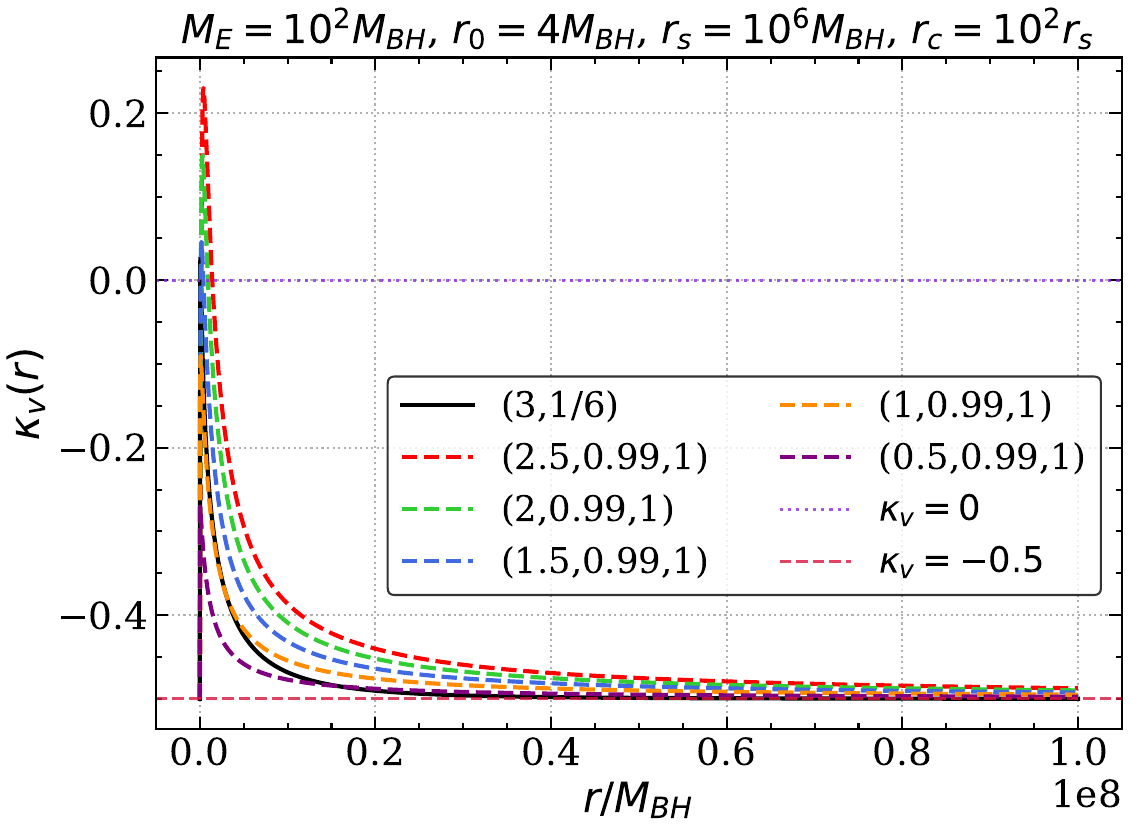}
\caption{The velocity power-law index $\kappa_v(r)$ as a function of radius $r \in [r_0,r_c]$. The zero-crossings and the asymptotic saturation to $\kappa_v \to (-1/2)$ are indicated by a violet dotted line and a crimson dashed line, respectively. Within the plotted DM region, $\kappa_v$ approaches closely but does not fully reach its asymptotic value of $(-1/2)$ for some cases.}\label{fig3}
\end{figure}

Now, to understand the physical content of $\{n_1,n_2,n_3\}$, let us construct the quantity
\begin{equation} \label{kappa}
    \kappa_X(r) := \frac{d\, \log X(r)}{d \log(r)}~,
\end{equation}
which probes the logarithmic slope (also referred to as the ``elasticity'') of a positive function $X(r)$ at a given radius $r$. Then, since the Newtonian circular velocity $v(r)$ depends on the enclosed mass $m(r)$ as $v^2(r)=m(r)/r$, we have $\kappa_m(r)=1+2\kappa_v(r)$ with $\kappa_m(r) = \kappa_{m_E}(r)\, \sigma(r)$, where $\sigma(r)$ is the fraction of the total mass contributed by the DM, i.e., $\sigma(r) = m_E(r)/m(r)$. After some straightforward algebra, we obtain an analytic expression (a similar expression can be found for the limiting ${}_1F_1$-based mass model as well):
\begin{widetext}
\begin{equation} \label{kappa2}
    \begin{aligned}
        \kappa_{m_E}(r) = \frac{r_0(1+\alpha)}{r-r_0}\, +\, n_1\, -\,\frac{n_1 (n_1+n_2)}{(n_1+n_3)}\left(\frac{r}{r_s}\right)^{n_3}\, \frac{{}_2F_1\left(\frac{n_1+n_3}{n_3},\frac{n_1+n_2+n_3}{n_3},\frac{n_1+2n_3}{n_3};-\left(\frac{r}{r_s}\right)^{n_3}\right)}{{}_2F_1\left(\frac{n_1}{n_3},\frac{n_1+n_2}{n_3},\frac{n_1+n_3}{n_3};-\left(\frac{r}{r_s}\right)^{n_3}\right)}~.
    \end{aligned}
\end{equation}
\end{widetext}
Hence, by mapping out $v(r)$ using astrophysical observations of stars, planets, or gas dynamics at different radii, we can directly probe the parameters $n_i$ $(i=1,2,3)$. For instance, $\kappa_v = -(1/2)$ at $r_0$ as $m_E'(r_0)$ vanishes, after which it first increases to attain a maximum and then decreases to saturate asymptotically to $(-1/2)$ as $m_E(r)$ saturates to $M_E$. Hence, $\kappa_v(r)$ will always have an even number of zero-crossings within the DM region, as illustrated in \ref{fig3}.
For the specific parameter choices, we find two zero-crossings for the choices $n_1=\{2.5,2,1.5\}$ and no zero-crossing for $n_1=\{1,0.5\}$. 

By construction, locations where $\kappa_v(r)=0$ correspond to the local extrema of $g^{rr}(r)$, which may have interesting consequences for geodesic motion as the radial velocity (in fact, $\dot{r}^2$) scales with $g^{rr}$. Also, the location of the maximum of $\kappa_v(r)$, or equivalently the location where $\kappa_m(r)-\kappa_\rho(r)=3$, can be used as an indicator of the transition at $r=r_s$. Note that such a location will always exist within the halo if its outer cutoff radius $r_c$ is large enough to have $\kappa_\rho(r_c) < -3$.

In fact, for $r_c \to \infty$, one finds $\kappa_\rho(r\sim r_c) = -3-\rm{min}(n_2,1)$, i.e., a power-law suppression in density for the mass model given by \ref{mE2}. Therefore, to maintain a finite total DM mass $M_E$ for an arbitrary choice of $r_c \gg r_0$, we must choose $n_2>0$. In fact, for illustration purposes with $r_c \gg r_s$, like in \ref{fig2} and \ref{fig3}, we will usually set $n_2$ close to unity. For the limiting case of \ref{1F1}, we find for $r_c \to \infty$ that $\rho(r\sim r_c) \sim A\, r^{n_1-3}\, \exp[-d(r/r_s)^{n_3}]+B/r^4$, with some constants $\{A,\,B\}$. For the purpose of illustration in \ref{fig2} and \ref{fig3} with $r_c \gg r_s$, we have chosen $n_1=3$, which reduces the profile to a Sérsic-type form in the absence of inner cutoff $r_0$ (i.e., without the central BH)~\cite{Sersic1963}. These relations can be used to link the asymptotic density decay and mass saturation scaling.

Due to the appearance of the combination $(n_1+n_2)$, the mass profile as written in \ref{mE2} may make it less obvious that $\{n_1,n_2\}$ play distinct physical roles as discussed above, namely $n_1$ controls the inner ($r \ll r_s$) density slope, while $n_2$ dictates the asymptotic density slope. To make it more transparent, we may use the standard identity relating ${}_2F_1$ and the incomplete $\beta$-functions~\cite{abramowitz1964handbook,NIST:DLMF,Arfken2012MathematicalMF,book:91229966}. In fact, it turns out that the mass profile in \ref{mE2} can be equivalently represented by
\begin{equation*} \label{mEbeta}
    m_E(r) = \frac{n_1 M_E}{n_3 N} \left(1-\frac{r_0}{r}\right)^{1+\alpha} \beta\left[\frac{(r/r_s)^{n_3}}{1+(r/r_s)^{n_3}},\, \frac{n_1}{n_3},\, \frac{n_2}{n_3} \right], 
\end{equation*}
which makes their role apparent. However, we will keep referring to the mass profile in \ref{mE2} as the ${}_2F_1$-based model in this paper.

Finally, the parameter $n_3$ controls how rapidly the mass profile transitions between the inner and outer regimes due to its appearance as $(r/r_s)^{n_3}$, and therefore encodes the sharpness of the breaking of scale invariance. Physically, it determines the characteristic width of the transition region around $r_s$. Large values of $n_3$ correspond to a relatively sharp, almost step-like change in $m_E(r)$ between the inner and outer behaviors, while smaller values produce a more gradual and extended transition. In the context of DM halos, this can be interpreted as a measure of how efficiently different dynamical processes, such as phase mixing, violent relaxation, or tidal effects, redistribute mass across scales. A sharply defined transition (large $n_3$) may indicate a more abruptly truncated or strongly structured halo, whereas a smoother transition (small $n_3$) reflects a more gradual redistribution of matter and a broader crossover between regimes. Thus, $n_3$ provides a physically meaningful way to quantify the height/slope of the mass profile at $r_s$. Observations of galaxy rotation curves may probe both $r_s$ and $n_3$. We will typically set $n_3$ to unity.

\section{Observable Signatures}
With the mass-based parametrization given by \ref{mE2} in place, we now turn to its potential observational signatures. Since the DM mass profile $m_E(r)$ exhibits different radial scalings across various regimes, it is natural to examine a range of observables like QNMs, TLNs, and GW fluxes from EMRIs, that can probe distinct aspects of the spacetime geometry deviating from the vacuum case. In particular, QNMs provide a direct handle on the properties of the unstable photon sphere; static TLNs, which vanish for the vacuum case, probe DM-induced deformations encoded in the perturbative sector; and EMRI systems accumulate phase shifts over a large number of in-band orbital cycles, providing a precise probe of the mass distribution. 

\subsection{Gravitational axial QNMs}
Gravitational perturbations $g_{\mu \nu}=g_{\mu \nu}^{(0)}+h_{\mu \nu}$ with $|h_{\mu \nu}|\ll1$ of the static, spherically symmetric background $g_{\mu \nu}^{(0)}$ given by \ref{metric} sourced by anisotropic, spherically symmetric matter naturally decouple into axial (odd parity) and polar (even parity) sectors. Since the background carries no parity-odd degrees of freedom (e.g., spin, magnetic field, vorticity, etc.), the axial sector will not couple to matter perturbations ($\delta T_{\mu \nu}$). This leads to a particularly simple Schrödinger-like equation $\partial_{r_*}^2 \psi_{\text{ax}} + (\omega^2-V_{\text{ax}})\psi_{\text{ax}}=0$~\cite{Barausse:2014tra, Cardoso:2021wlq, Zhao:2023tyo,Speeney:2024mas, Chakraborty:2024gcr,Pezzella:2024tkf,Bhowmik:2026owi} with potential
\begin{equation} \label{axial}
    V_{\text{ax}}(r)=f(r) \left[\frac{\ell\, (\ell+1)}{r^2}  - \frac{6\, m(r)}{r^3} +\frac{m'(r)}{r^2} \right]~,
\end{equation}
where $\{f(r),m(r)\}$ are provided by \ref{metriccomp}, $\ell \geq 2$ is the angular multipole index, and $dr/dr_*=\sqrt{f(r)(1-2m(r)/r)}$ defines the tortoise coordinate $r_*$. 

Note that the potential $V_\text{ax}$ vanishes at the horizon because $f(r_H)=0$, and at spatial infinity due to asymptotic flatness. Thus, it must have at least one extremum outside $r_H$. We also recall that the corresponding vacuum potential $V_{\text{ax,vac}}$ has exactly one maximum outside the horizon. In fact, for realistic dilute $M_E/r_s \ll 1$ DM halos, we will still have a single maximum in the potential $V_\text{ax}$. Then, as expected from the eikonal correspondence for large $\ell$'s (which is still quite accurate even for lower values, like $\ell=2$), the location of this maximum will almost coincide with that of the photon sphere, $r_\gamma$, specified by the roots of $\Delta(r):= r f'(r)-2 f(r)=0$.

Due to asymptotic flatness, $\Delta(r)$ admits an odd (counted with multiplicity) number of roots, or light rings, in the domain of outer communication~\cite{Ghosh:2021txu, Ghosh:2023kge}. In particular, considering $r_0=4M_{BH}$, we still get $r_\gamma=3M_{BH}$ as a valid solution, as the region inside $[r_H,r_0]$ is devoid of any DM. Hence, we must have an even number of light rings inside the halo region, where $\Delta(r)$ can be simplified using the Einstein's equations in \ref{finalEin} as
\begin{equation} \label{LR}
    \Delta(r) = - \frac{2\, f(r)\, [r-3\, m(r)]}{r-2\, m(r)}=\frac{2\, f(r)\, T(r)}{\rho(r)}~,
\end{equation}
which is valid everywhere and follows as a special case of the more general derivation in Ref.~\cite{Ghosh:2023kge}, and $T = -\rho+2p_t$ is the trace of the matter stress-energy tensor $(T^E)^{\mu}_{\nu}$. Therefore, the locations $r_\gamma$ of light rings inside the DM halo are characterized by the positions where the matter is locally traceless, i.e., $T(r_\gamma)=0$, or the spacetime is locally Ricci-scalar-flat, i.e. $R(r_\gamma)=0$. For realistic dilute DM halos, however, $T(r>r_0)$ will always have the same sign, and as a result there will be no other light ring. In fact, it follows from \ref{LR} that $T(r>r_0)$ is strictly negative for DM profiles whose enclosed mass function $m(r)$ never exceeds $r/3$.
\begin{figure}[h!]
\centering
\includegraphics[width=\linewidth]{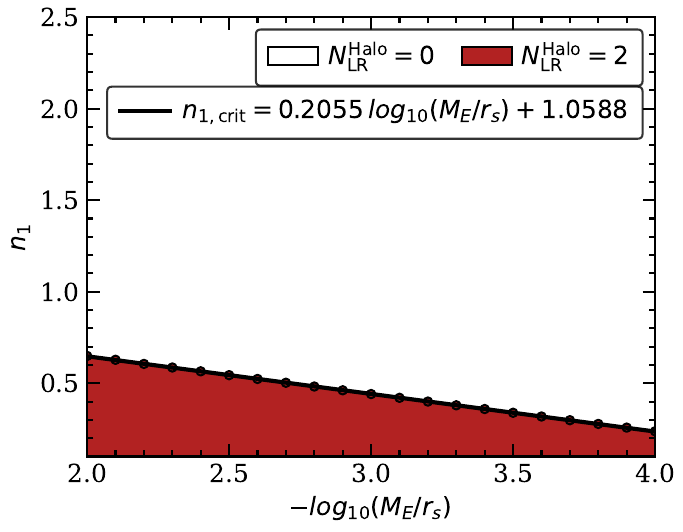}
\caption{Plot showing the variation in the number $N_{\rm LR}^{\rm Halo}$ of light rings within the halo modeled by the mass profile given by \ref{mE2}, with compactness $z=M_E/r_s \in [10^{-4},10^{-2}]$ (with $r_s$ fixed at a radius $10^6$ and $M_E$ varied accordingly in units of $M_{BH}=1$) and $n_1 \in [0.1,2.5]$. All other parameters are fixed as mentioned in the caption of \ref{fig2}. The solid black line represents critical values of $n_{1,\, \rm crit}$ linearly fitted against $\log_{10}z$. This line divides the $(n_1,z)$-parameter space into two regions: the white one, with no light ring inside the halo, and the red one, with two light rings inside the halo.}\label{lrcount}
\end{figure}

For the fixed halo parameters specified in the caption of \ref{fig2}, our numerical analysis indicates that the ${}_2F_1$-based model admits an even (counted with multiplicity) number of light rings within the halo only for $n_1 \lesssim 0.2382$. However, such small values of $n_1$ lead to an unrealistically rapid radial growth of DM mass in the inner region ($r_0 \ll r \ll r_s$), which may render the profile unphysical. In particular, our construction is built upon the Einstein-cluster-like configuration, for which the radial pressure is taken to be negligibly small (i.e., $p_r \approx 0$) and the halo support is predominantly provided by the tangential motion of the DM particles. A realistic collisionless halo may not satisfy $p_r \approx 0$ exactly, and a small but finite radial velocity dispersion ($\sigma_r$) can therefore be used to quantify departures from the aforementioned ideal setup. 

As a simple consistency check, we consider the Newtonian Jeans equation~\cite{j_binney_galactic_1987, mo2010galaxy} in the region $r \gg r_H$, where general relativistic corrections are small. In the isotropic limit~\cite{j_binney_galactic_1987, mo2010galaxy}, we obtain
\begin{equation} \label{sigmar}
    \sigma_r^2 (r \gg r_H) \approx \frac{1}{\rho(r)}\int_{r}^{r_c} \frac{\rho(\tilde r)\, m(\tilde r)}{\tilde r^2}d\tilde r,
\end{equation} 
as required for the dynamical equilibrium of the system. Then, a straightforward calculation shows that in the inner halo region ($r_0 \ll r \ll r_s$), we have $\sigma_r^2 \sim (r/r_s)^{n_1-1}$ for $n_1 < 2$, and $\sigma_r^2 \sim (r/r_s)\, \log(r_s/r)$ for $n_1=2$, as the leading DM contribution.  This implies a steep inward growth of the velocity dispersion for $0<n_1 \ll 1$, indicating that a very steep inner mass profile with a small radial pressure may require an excessively large kinetic support to maintain equilibrium, thereby making the Einstein-cluster-like configuration progressively less reliable. Such behavior is also unlikely to be compatible with a physically viable steady-state distribution function.\footnote{Besides causing large velocity dispersion, a steeper inner slope (i.e., low values of $n_1$) would lead to a significantly enhanced annihilation luminosity $L_{DM} \sim 4\pi \int r^2 \rho^2(r) dr$ for generic weakly interacting DM candidates. Such a large annihilation rate would efficiently deplete the central DM spike, leading to the formation of an annihilation plateau.}

These considerations suggest that very low values of $n_1$ may not be conducive to a relaxed halo. Hence, for illustration purposes, we restrict our analysis to $n_1 \geq 0.5$ for the halo compactness $z:= M_E/r_s = 10^{-4}$. For other values of $z$, we always consider those $n_1$ that lead to no additional light ring (i.e., $r > 3m(r)$ holds true) inside the halo: see \ref{lrcount} for an illustrative plot. As an example, for $z=10^{-2}$, we restrict ourselves to $n_1 \geq 0.7$, to satisfy the above consideration.
\begin{figure*}
    \centering
    \includegraphics[width=0.49\textwidth]{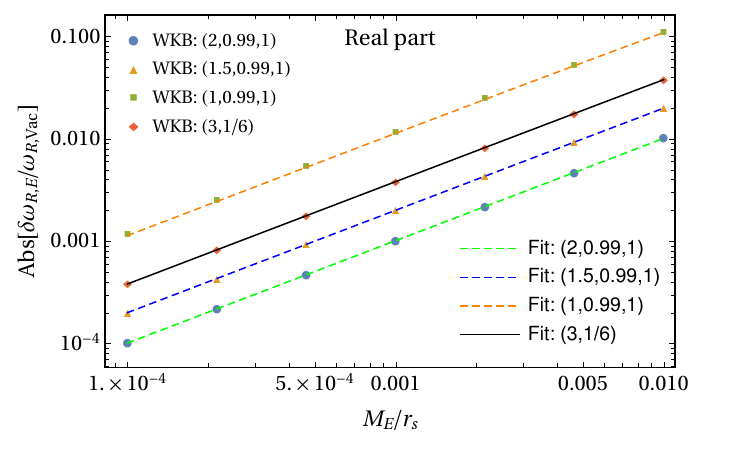}
    \includegraphics[width=0.49\textwidth]{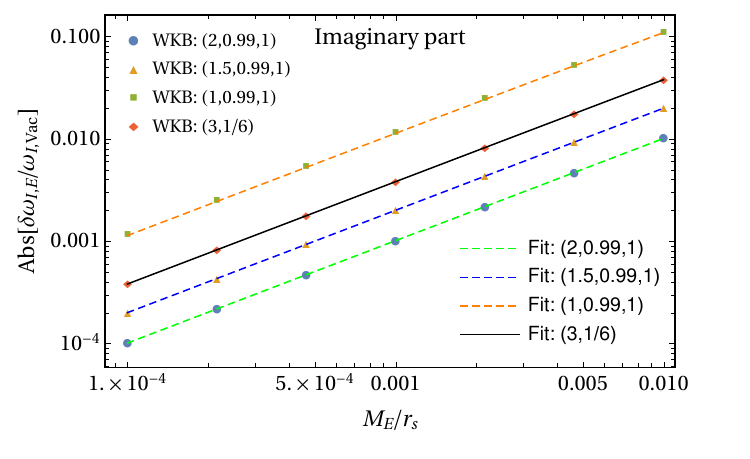}
    \caption{
    Left:  Variation in the relative shift in the real part of $\ell=2$ gravitational axial QNMs with respect to compactness in log-log scale. 
    Right: Variation in the relative shift in the imaginary part of $\ell=2$ gravitational axial QNMs in log-log scale. The plot markers represent the numerically obtained values computed using the 6th-order WKB method, whereas the lines represent the corresponding straight-line fits. Both the real and imaginary parts lead to the same relative differences for a given compactness. Moreover, the actual slopes (not in log-log scale) of these lines from bottom to top (i.e., green to orange) are $\{1.0106, 2.0028, 3.8165, 11.3584\}$, which match closely with those calculated using the analytical formula in \ref{slope}.
    }
    \label{fig4}
\end{figure*}

Now, to understand how the DM halo affects the dominant ($\ell=2$) axial QNMs for the fundamental mode, we analyze them using the 6th-order Wentzel-Kramers-Brillouin (WKB) method. 
A few illustrative cases are shown in \ref{fig4}, where we consider an inner cutoff radius $r_0=4M_{BH}$, transition scale $r_s=10^6M_{BH}$, outer cutoff radius $r_c=10^8M_{BH}$, and compactness $z \in [10^{-4},10^{-2}]$. Moreover, we have chosen $\alpha=1$, $n_2=0.99$, $n_3=1$ and we consider $n_1$ from the list $n_1 \in \{1,1.5,2\}$ for the dashed lines, while we fix $(n_1=3,d=18,n_3=1/6)$ for the solid black line representing an illustrative case of \ref{1F1}. For these choices of parameters, the spacetime has a single light ring at $r=3M_{BH}$.

Interestingly, in this WKB analysis, both the real ($\delta \omega^E_R/\omega_{R}^{\text{vac}}$) and imaginary ($\delta \omega^E_I/\omega_{I}^{\text{vac}}$) parts of the QNM frequencies show the same relative difference with respect to the corresponding vacuum values for any value of the compactness parameter $z=M_E/r_s$. This is expected for the following reason. Since $r_0=4M_{BH}$ is taken to be bigger than the light ring radius $r_\gamma=3M_{BH}$, the eikonal axial potential takes the form
\begin{equation} \label{eikaxi}
    V_{\text{ax,eik}}(r) \approx \frac{\ell^2\, f_s(r)}{r^2}\, e^{\mu_0} = V_{\text{ax,eik}}^{\text{vac}}(r)\, e^{\mu_0}~,
\end{equation}
by ``zooming-in'' the general potential in \ref{axial} near the light ring.
\begin{figure}[h!]
\centering
\includegraphics[width=\linewidth]{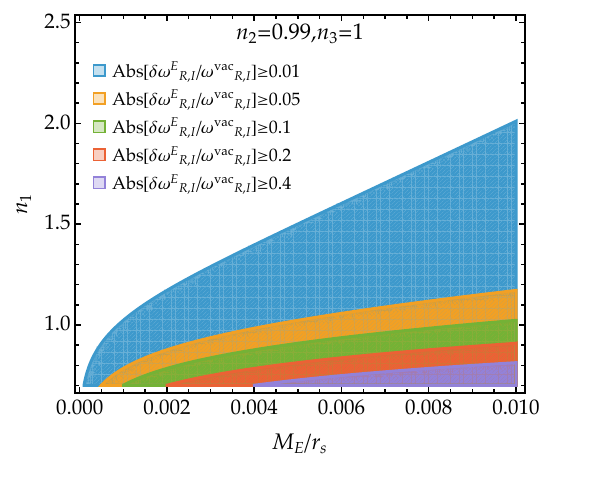}
\caption{Variations of the relative difference in $\ell=2$ axial gravitational QNMs (computed using the 6th-order WKB method) with $n_1 \in [0.7,2.5]$ and compactness $z \in [10^{-4},10^{-2}]$ for our ${}_2F_1$-based model. Lower values of $n_1$ produce larger relative shifts in the real part of the QNM frequencies. This can be used to put lower limits on the allowed values of $n_1$.}\label{fig5}
\end{figure}
Then, a straightforward application of the eikonal correspondence implies that
\begin{equation} \label{RI}
    \frac{|\delta\omega^E_{R,I}|}{|\omega_{R,I}^{\text{vac}}|}\Bigg{\rvert}_{\text{axial}}\overset{(\text{eik})}{=} 1-e^{\mu_0/2} \approx \frac{|\mu_0|}{2}~.
\end{equation}
Hence, in the eikonal limit, the DM halo produces an overall redshift in the vacuum QNMs~\cite{Cardoso:2021wlq}. Moreover, observational measurements of these shifts provide direct access to the horizon surface gravity $\kappa_H =e^{\mu_0/2}/(2r_H)$. Now, using \ref{finalEin}, one further obtains the slope of the relative shift in QNMs with respect to the compactness $z=M_E/r_s$ as
\begin{equation}\label{slope}
    \text{Slope}\overset{(\text{eik})}{:=} \frac{d}{dz}\left(\frac{|\delta\omega^E_{R,I}|}{|\omega_{R,I}^{\text{vac}}|}\right) \approx \frac{r_s}{r_c}+\frac{r_s}{M_E} \int_{r_0}^{r_c} \frac{m_E(r)\, dr}{(r-r_H)^2}~,
\end{equation}
a universal relation that is independent of the specific form of the mass/density model. Here, to obtain the final expression, we have fixed $r_s$, and vary $M_E=z\, r_s$ accordingly in the mass profile $m_E(r)$ given by \ref{mE2}. Note that the slope defined above is independent of the total DM mass $M_E$ at leading order in $z$. 

From \ref{fig4}, it is clear that for fixed $n_2$ and $n_3$, the slope is highly sensitive to the choice of $n_1$, providing a direct probe to observational constraints on its value. In particular, for fixed values of $n_2=0.99$, $n_3=1$ and other model parameters, \ref{fig5} illustrates this dependency with $\{n_1,z\}$. Since waveform mismatches relevant for detectability scale as $\sim (\text{signal-to-noise ratio})^{-2}$, future detectors might be sensitive to probe effects associated with $n_1$. However, robust constraints will require a careful treatment of parameter degeneracies within the full model, a study that we defer to future work.

\begin{figure}[h!]
\centering
\includegraphics[width=\linewidth]{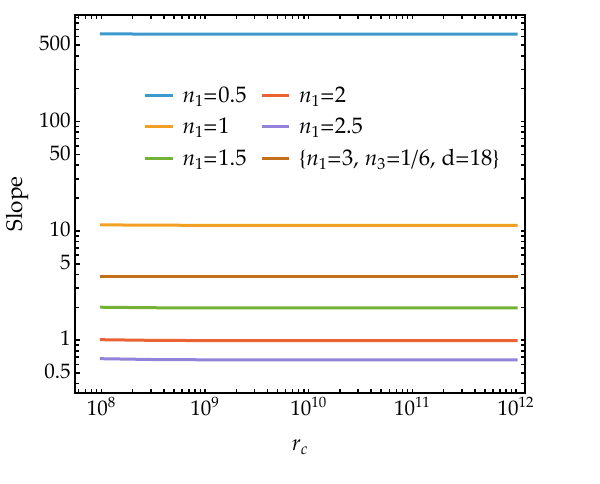}
\caption{Log-log plot showing the variations of the QNM slope as defined in \ref{slope} with the halo outer cutoff radius $r_c \in [10^{8},10^{12}]$, and $n_1 \in \{0.5,1,1.5,2,2.5\}$ for the ${}_2F_1$-model ($n_2=0.99$ and $n_3=1$) and for a limiting case of the ${}_1F_1$-model. The rest of the halo parameters are fixed as mentioned in the caption of \ref{fig2}.}\label{qnmvsrc}
\end{figure}

Moreover, \ref{slope} suggests that the slope is rather insensitive to the outer cutoff radius $r_c$. This is because $r_s/r_c \ll 1$, and the integral kernel in \ref{slope} near $r_c$ is suppressed quadratically in radius. As a result, the slope 
decays very weakly with increasing $r_c$ when all other parameters are kept fixed. This is illustrated in \ref{qnmvsrc}. We will come back to this important observation at the end of the next subsection on static TLN.
\subsection{Gravitational axial TLN: Static case}
Let us now turn to the gravitational tidal response and the associated TLNs~\cite{Hinderer:2007mb,  Binnington:2009bb, Damour:2009vw} of these BHs. For simplicity, let us focus again on the axial case. It is well known that the static ($\omega=0$) TLN of a vacuum Schwarzschild BH vanishes~\cite{Damour:2009vw, Binnington:2009bb, Kol:2011vg, 2013arXiv1304.2228C,  Gurlebeck:2015xpa, Chia:2020yla, Bhatt:2023zsy, LeTiec:2020bos, Charalambous:2021mea, Chakraborty:2026qru, hui2022ladder-678, achour2022hidden-8c7, charalambous2021hidden-5e0, ivanov2023vanishing-9aa, creci2021tidal-42e, Charalambous:2021mea, Singha:2025xah,Rodriguez:2026iot}. This follows from the fact that the corresponding perturbation equation admits no solution that is simultaneously regular at the horizon and decays at spatial infinity. In the presence of the DM halo, however, this property is generally lost. We will indeed see that the environment modifies the perturbation equation in such a way that a horizon-regular solution can acquire a non-trivial decaying component at infinity, thereby giving rise to a non-zero static TLN~\cite{Chakraborty:2024gcr,  Cardoso:2019upw, Cardoso:2021wlq, DOnofrio:2026ulh, Zhao:2026eti,Chakraborty:2026qru,Bhowmik:2026owi}. Consequently, the tidal response provides a novel observational probe for distinguishing an isolated vacuum BH from one embedded in a DM environment.

The governing static axial perturbation equation is  $(\partial_{r_*}^2 -V_{\text ax})\, \psi_{\text{ax}}=0$, with the potential given by \ref{axial}. The corresponding vacuum system, i.e., $m(r)=M_{BH}$, admits two algebraically independent solutions for $\ell=2$:
\begin{equation} \label{sols}
    \begin{split}
        &\psi^{\text{ax}}_{\text{reg}}(r) = r^3, \, \, \\
        &\psi^{\text{ax}}_{\text{irr}}(r)=r^3 \log[f_s(r)] \\
        &\kern4em 
        +\frac{r_H}{12\, r} \left(12\, r^3 + 6\, r^2\, r_H + 4\, r\, r_H^2 + 3\, r_H^3 \right)~,
    \end{split}
\end{equation}
with a Wronskian $W(r) = r_H^5/f_s(r)$. The former solution $\psi^{\text{ax}}_{\text{reg}}$ is globally regular and can be identified as the tidal field asymptotically, where it grows as $r^{\ell+1}$. The second solution $\psi^{\text{ax}}_{\text{irr}}$ has a logarithmic divergence at the horizon and can be identified as the tidal response decaying as $r^{-\ell}$ at spatial infinity. As a result, the allowed complete solution $\psi_{\text{ax}}=c_{\text{reg}}\, \psi^{\text{ax}}_{\text{reg}} +c_{\text{irr}}\, \psi^{\text{ax}}_{\text{irr}}$ (where the $c$'s are arbitrary constants) must have $c_{\text{irr}}=0$, leading to the vanishing of the (dimensionless) TLN defined as~\cite{Hui:2020xxx, Singha:2025xah,Chakraborty:2026qru}
\begin{equation} \label{TLN}
    \Lambda^{(\ell)}_{\text ax} = - \frac{(\ell-1)}{(\ell+2)} \times \frac{1}{\mathcal{L}^{2\ell+1}}\left(\frac{\text{Coeff. of}\, \,  r^{-\ell}\, \, \text{in}\, \, \psi^{\text{ax}}_\infty}{\text{Coeff. of}\, \,  r^{\ell+1}\, \, \text{in}\, \, \psi^{\text{ax}}_\infty}\right)~,
\end{equation}
where $\mathcal{L}$ is a characteristic length scale of the system and $\psi^{\text{ax}}_\infty$ is a shorthand for $\psi_{\text{ax}}(r)$ at $r \to \infty$. 

Now, in the presence of a DM halo, the situation is starkly different. In fact, the perturbation equation no longer admits an analytical solution in general, although the corresponding TLN can still be computed numerically. Before describing the numerical procedure, however, we first develop a perturbative analytical treatment that will turn out to agree remarkably well with the full numerical result. 

To this end, we expand all relevant quantities $X(r)$, such as $\{\psi_{\text{ax}}(r), V_{\text ax}(r)\}$, appearing in the perturbation equation as $X_0(r) + \epsilon\, X_1(r)$, where $X_0(r)$ denotes the vacuum value, $\epsilon$ is a bookkeeping expansion parameter, and $X_1(r)$ represents the leading correction induced by the DM halo. The parameter $\epsilon$ may be regarded as arising from the replacement $M_E \to \epsilon\, M_E$, so that $\epsilon=0$ corresponds to the vacuum case and $\epsilon=1$ recovers the full DM configuration. Note also that for a realistic diffused halo, such expansions are allowed as $|X_1(r)| \ll |X_0(r)|$.  In particular, the quantities $\mu_0$ and $\mu(r)$ enter the perturbation equation through the factors $e^{\mu_0}$ and $e^{\mu(r)}$ appearing in $f(r)$, as defined in \ref{metriccomp}. Under the scaling $M_E \to \epsilon\, M_E$, they transform to $e^{\epsilon \, \mu_0}$ and $e^{\epsilon \, \mu(r)}$, respectively. Since $|\mu(r)|$ remains small throughout the domain of interest, we consistently linearize it as $1+\epsilon\, \mu(r)$. Indeed, $|\mu(r)|$ is bounded in the entire interval within $[2M_E/r_c \ll 1,|\mu_0|\ll 1]$, ensuring the validity of the approximation.
\begin{figure}[h!]
\centering
\includegraphics[width=\linewidth]{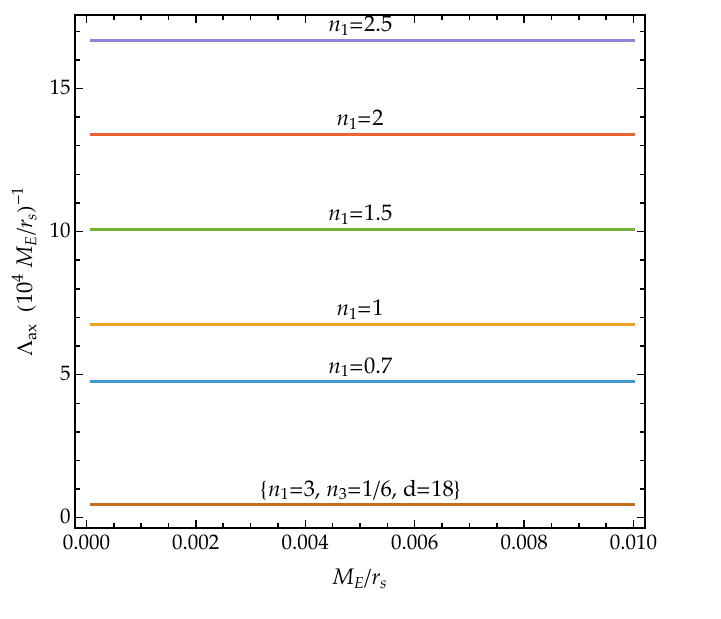}
\caption{Plot showing the variations of the TLN computed using \ref{fTLN} with the halo compactness $z \in [10^{-4},10^{-2}]$, and $n_1 \in \{0.7,1,1.5,2,2.5\}$ for the ${}_2F_1$-model ($n_2=0.99$ and $n_3=1$) and the bottom line for the limiting case of ${}_1F_1$-model. The lines are horizontal, indicating that $(\Lambda_{\text ax}/z)$ is independent of compactness.}\label{fig6}
\end{figure}

\begin{table*}[!t] 
\centering
\begin{tabular}{c c c c c c c}
\hline\hline
$n_1$ & $0.5$ & $1$ & $1.5$ & $2$ & $2.5$ & $\{n_1=3, n_3=1/6, d=18\}$
\\
\hline
Coeff. of $r^3\, \, \text{in}\, \, \psi_\infty$
& 0.131396
& 0.125113
& 0.125020
& 0.125010
& 0.125007 
& 0.125038
\\
Coeff. of $r^{-2}\, \, \text{in}\, \, \psi_\infty$
& $-1.787\times10^{30}$
& $-3.383\times10^{30}$
& $-5.045\times10^{30}$
& $-6.697\times10^{30}$
& $-8.337\times10^{30}$
& $-2.2785\times10^{29}$
\\
TLN (Numerical)
& 3.40088
& 6.75933
& 10.08797
& 13.39200
& 16.67327
& 0.455561
\\
TLN [using \ref{fTLN}]
& 3.39988
& 6.75871
& 10.08756
& 13.39124
& 16.67243 
& 0.454558
\\
\hline\hline
\end{tabular}
\caption{Asymptotic coefficients of the static axial perturbation and the corresponding TLN ($\ell=2$) for several values of the halo parameter $n_1 \in \{0.5,1,1.5,2,2.5\}$ for the ${}_2F_1$-model ($n_2=0.99$ and $n_3=1$). The last column refers to the limiting case of the ${}_1F_1$-model. We have fixed $r_0=4M_{BH}$, $M_E=100M_{BH}$, $r_s=10^6M_{BH}$, $r_c=10^2 r_s$, $\alpha=1$. The theoretical TLNs computed using \ref{fTLN} agree quite well with the numerical values obtained by directly solving the perturbation equation at $r>r_c$.}
\label{tab:TLNcomparison}
\end{table*}

Then, up to linear order in $\epsilon$, the perturbation equation breaks into two coupled second-order systems: $\mathcal{D}^{\text{ax}}_0\, \psi^{\text{ax}}_0 =0$ and $\mathcal{D}^{\text{ax}}_0\, \psi_1 \approx \mathcal{S}[\psi^{\text{ax}}_0]$, with
\begin{equation} \label{D0S}
    \begin{split}
        &\mathcal{D}^{\text{ax}}_0 \equiv r^3\, f_s(r)\, \partial_r^2+ r\, r_H\, \partial_r - 3\, (2\, r-r_H)~,\\
        & \mathcal{S}[\psi^{\text{ax}}_0] \equiv 16\, \pi\, r^6\, \rho(r)~,
    \end{split}
\end{equation}
where we have replaced $m_E'(r)=4\, \pi\, r^2\, \rho(r)$ using one of the Einstein's equations in \ref{finalEin} and this quantity can be computed using the mass model given by \ref{mE2}. Demanding vacuum regularity, we have also identified $\psi^{\text{ax}}_0(r) \sim \psi^{\text{ax}}_{\text{reg}}(r)$ up to an overall constant. The second source equation can be solved by the ``variation of parameters'' method~\cite{book:91852226} as
\begin{equation} \label{psi1}
    \begin{split}
        &\psi^{\text{ax}}_1(r) \approx \frac{16\, \pi}{r_H^5} \Bigg[\psi^{\text{ax}}_{\text{irr}}(r) \int_{r_H}^{r} y^3\, \rho(y)\, \psi^{\text{ax}}_{\text{reg}}(y)\, dy  \\
        &\kern6em +\, \psi^{\text{ax}}_{\text{reg}}(r) \int_{r}^{\infty} y^3\, \rho(y)\, \psi^{\text{ax}}_{\text{irr}}(y)\, dy\Bigg]~.
    \end{split}
\end{equation}
Now, for a DM halo with a large but finite outer cutoff radius at $r_c \gg r_s$, the contribution from the second integral vanishes for the field $\psi^{\text{ax}}_1(r \to \infty)$ measured outside the extension of the DM halo ($r>r_c$), while the first integral gives a non-zero value only in $[r_0,r_c]$. Finally, using the background homogeneous solutions in \ref{sols} and the definition of TLN in \ref{TLN}, we obtain for $\ell=2$:
\begin{equation} \label{fTLN}
    \Lambda_{\text ax} \approx \frac{\mathcal{M}_6[\rho]}{5\, r_s^5}~,
\end{equation}
where $\mathcal{M}_k[X] \equiv 4\, \pi\, \int_{r_0}^{r_c} r^k\, X(r)\, dr$ is the $k$-th radial moment of $X(r)$ in the interval $[r_0,r_c]$. For example, the $k=2$ case gives the total halo mass $M_E$. Note that $\Lambda_{\text ax}$ is proportional to the sixth radial moment of the DM density $\rho(r)$. In turn, $\Lambda_{\text ax}$ is proportional to the total halo mass $M_E$, which vanishes in the vacuum (Schwarzschild) case. We have also identified $\mathcal{L}$ with the transition radius $r_s$, as it is the natural scale associated with the DM halo~\cite{Chakraborty:2024gcr,Chakraborty:2026qru}. The above equation is universal, independent of the specific form of the mass/density model.  

However, in the presence of a DM halo, the TLN of the whole system is indeed non-zero. A few illustrative cases are shown in \ref{fig6} with fixed $r_0=4M_{BH}$, $M_E=100M_{BH}$, $r_s=10^6M_{BH}$, $r_c=10^2 r_s$, $\alpha=1$ for all the cases, and $\{n_2=0.99,\, n_3=1\}$ for the ${}_2F_1$-based models. In fact, a careful calculation shows that the TLN scales as $\Lambda_{\text ax} \sim [z\, \kappa_{m_E}(r_c)/25]\,(r_c/r_s)^4$ up to an overall $\mathcal{O}(1)$ multiplicity, where $z=M_E/r_s$ is the compactness of the DM halo and the quantity $\kappa_{m_E}$ is defined in \ref{kappa2}. Thus, $\Lambda_{\text ax} \propto z$ as supported by \ref{fig6} also (contrast it with the corresponding QNM case in \ref{fig4}).

\begin{figure}[h!]
\centering
\includegraphics[width=\linewidth]{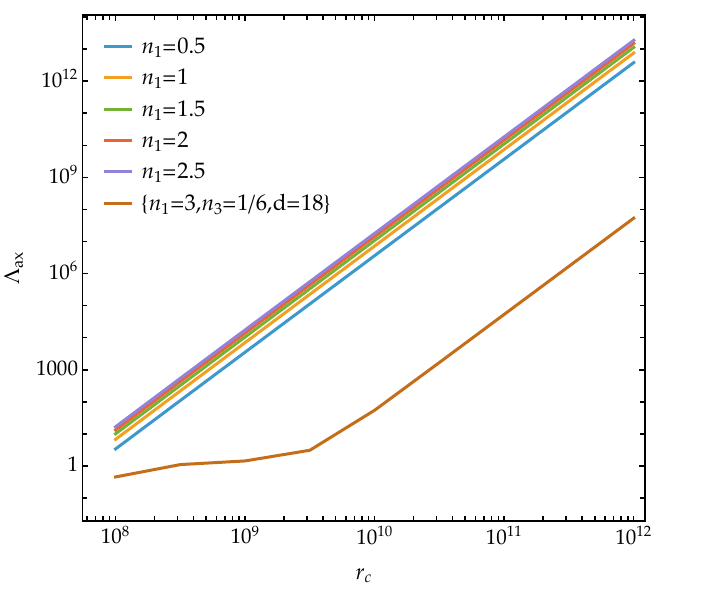}
\caption{Log-log plot showing the variations of the TLN computed using \ref{fTLN} with $r_c \in [10^{8},10^{12}]$, and $n_1 \in \{0.5,1,1.5,2,2.5\}$ for the ${}_2F_1$-model ($n_2=0.99$ and $n_3=1$) and a limiting case of ${}_1F_1$-model. The rest of the halo parameters are fixed as mentioned in the caption of \ref{tab:TLNcomparison}.}\label{fig7}
\end{figure}

Since $\kappa_{m_E}(r_c)$ decreases sharply with $r_c$ when the rest of the parameters are fixed, the overall growth of the TLN with the outer cutoff radius is milder than $r_c^4$. Nevertheless, $\Lambda_{\text ax}$ increases with $r_c$, as highlighted in \ref{fig7} (contrast it with the corresponding QNM case in \ref{qnmvsrc}). This is expected, as a more diffuse halo is more easily deformed by an external perturbation. For example, our ${}_2F_1$-based model leads to a $\Lambda_{\text ax}$ scaling as $\Lambda_{\text ax}\propto r_c^{4-\text{min}(n_2,1)}$ for all $n_2>0$, which increases with $r_c$ unless the DM density decays so rapidly that the outer halo region fails to contribute significantly. For the limiting ${}_1F_1$-model this automatically reduces to $\Lambda_{\text ax}\propto r_c^3$. The change of slope around $r_c\sim 10^9$ in \ref{fig7} highlights the crossover radius beyond which the model runs parallel to the ${}_2F_1$ curves. Interestingly, the TLNs for the ${}_2F_1$-based model seem to follow a simple relation: $\Lambda_{\text ax} \approx 10^4\, z\,(0.1191 + 6.6340\, n_1)$ as we vary $\{z,n_1\}$ while other DM parameters are fixed, as quoted in the caption of \ref{tab:TLNcomparison}
for a compactness $z = 10^{-4}$. We compare theoretically obtained values using \ref{fTLN} with those obtained via a numerical method, which directly integrates the governing static axial perturbation equation $(\partial_{r_*}^2 -V_{\text ax})\, \psi_{\text{ax}}=0$ in $[r_{\text{min}}=r_H+\delta, r_{\text{out}} > r_c]$ with $0 \lesssim \delta \ll 1$ to obtain the field $\psi_{\text{ax}}(r_{\text{out}})$. Then, one can easily extract the coefficients of $r^3$ and $r^{-2}$ in $\psi_{\text{ax}}(r_{\text{out}})$, and deduce the TLN using \ref{TLN} for $\ell=2$ and $\mathcal{L}=r_s$. For this numerical method, one needs two initial conditions at $r=r_{\text{min}} \approx r_H$, namely $\psi_{\text{ax}}(r_{\text{min}})=1$ and $\psi_{\text{ax}}'(r_{\text{min}}) \approx [\ell(\ell+1)+1-s^2]/r_H$. Here, we obtain the second condition by solving the perturbation equation at the horizon and setting the coefficient of the divergent piece to zero.

Let us complete this subsection with an interesting observation. QNMs and static TLNs seem to be most sensitive to two complementary regions of the DM halo. In particular, QNMs are predominantly sensitive to the inner regions of the halo, because the integral kernel in \ref{slope} falls off as $r^{-2}$. In contrast, \ref{fTLN} indicates that the static TLNs primarily probe the outer region of the halo, owing to the integral kernel in \ref{fTLN} growing as $r^6$. This complementarity offers a useful means of probing both the inner and outer halo structures, thereby providing a more complete characterization of the halo profile than either observable alone.

In fact, the sensitivity of TLNs to the outer halo regions is intimately connected to the absence of an intrinsic length scale in static ($\omega=0$) perturbations. Unlike QNMs, whose finite frequency localizes the wave dynamics near the potential barrier around the photon sphere, static perturbations efficiently accumulate contributions from the entire matter distribution. This accumulation becomes particularly sensitive at large radii due to the $r^{2(\ell+1)}$-weighting of the density profile in \ref{fTLN}. In turn, it also causes $\Lambda_{\text ax}$ to increase with the outer cutoff radius $r_c$ as discussed earlier, potentially rendering the tidal response measurable by GW observations and offering a clear signature to distinguish a DM environment from exotic compact objects. 

\subsection{EMRI fluxes}
To complete our discussion of the parameterized DM halo model and of its observable signatures, here we compute GW fluxes generated by an EMRI in the DM halo. Our model for the parameterized DM halo follows the Einstein cluster prescription of \ref{sec:model_construction}. As such, we treat perturbations according to the framework established in Ref.~\cite{Cardoso:2022whc}, which derived a set of equations governing the gravitational and environmental fluctuations induced by the orbiting particle. Here, we present some essential elements of the computation. Further details of the perturbative treatment and numerical flux calculations can be found in Ref.~\cite{Cardoso:2022whc} and in related works, like Refs.~\cite{Figueiredo:2023gas,Speeney:2024mas,Gliorio:2025cbh}.

Following closely the perturbative treatments of Regge-Wheeler and Zerilli, gravitational and environmental perturbations may be split into axial/odd and polar/even sectors based on their behavior under parity transformations. In the axial sector, gravitational and environmental perturbations decouple; the dynamics are governed entirely by a single Schrödinger-like master equation for the function $\psi_{\text{ax}}$, which reads 
\begin{equation}
\partial_{r_*}^2 \psi_{\text{ax}} + (\omega^2-V_{\text{ax}}) \psi_{\text{ax}}=S_{\text{ax}}.
\end{equation}
The potential $V_{\text{ax}}$ and the tortoise coordinate $r_*$ are defined in the same manner as~\ref{axial}. The term $S_{\text{ax}}$ is sourced by the secondary point mass $m_p$ (with mass ratio $q \equiv m_p/M_{BH} \ll 1$), undergoing circular motion with orbital radius $r_p$ and angular frequency $\Omega_p$. The explicit form of $S_{\text{ax}}$ can be found in the Appendix of Ref.~\cite{Cardoso:2022whc}.

The polar sector does not admit such a simple master equation, instead requiring the evaluation of five coupled first-order ODEs for the gravitational fluctuations $\{H_0,H_1,K\}$ and the environmental fluctuations $\{W,\delta\rho\}$. Here, $\delta\rho$ and $W$ are related to the fluid's density and velocity perturbations, respectively. The system of first-order equations can be written as
\begin{equation}
\label{eq:polar_pert}
\partial_r\vec{\psi}_{\text{pol}}=\boldsymbol{A}\vec{\psi}_{\text{pol}}+\vec{S}_{\text{pol}},
\end{equation}
where the vector $\vec{\psi}_{\text{pol}}=(H_0,H_1,K,W,\delta\rho)$. Once again, explicit expressions for the polar source terms $\vec{S}_{\text{pol}}$ and the matrix $\boldsymbol{A}$ can be found in the Appendix of Ref.~\cite{Cardoso:2022whc}.

Due to the decoupling of the odd-parity gravitational and environmental perturbations, the non-vacuum GW fluxes in the axial sector are well described by a simple rescaling of the vacuum quantities by a redshift factor $\gamma=1-z$, for realistic values of $z$, as~\cite{Cardoso:2022whc,Figueiredo:2023gas}
\begin{equation}
(\Omega^{\text{vac}},\, \omega^{\text{vac}},\, m_p)\rightarrow\left(\frac{\Omega}{\gamma},\, \frac{\omega}{\gamma},\, \gamma\, m_p\right).
\end{equation}
The compactness $z = M_E/r_s$ is expected to be very small in realistic galactic halos, and therefore DM-induced effects in the axial sector are highly suppressed: for Milky-Way like parameters of $M_E=10^{12}M_\odot$ and $r_s=20 \text{ kpc}$, the compactness $z\sim10^{-7}$. 
Therefore, we choose to focus on the polar sector, where the coupling between gravitational and fluid modes becomes non-trivial. 

We solve the system of ODEs in \ref{eq:polar_pert} via a shooting method. Following Ref.~\cite{Speeney:2024mas}, we first assume ingoing and outgoing boundary conditions for the gravitational fields $\{H_0,H_1,K\}$ at the horizon as 
\begin{equation}
\begin{aligned}
\label{eq:Hor_BCs}
H_0(r\rightarrow 2 M_{BH})&=e^{-i\omega r_*}\sum_{i=-1}^{n^H_{\text{max}}}b_i^{(H_0)}(r-2M_{BH})^i\\
H_1(r\rightarrow2M_{BH})&=e^{-i\omega r_*}\sum_{i=-1}^{n^H_{\text{max}}}b_i^{(H_1)}(r-2M_{BH})^i\\
K(r\rightarrow2M_{BH})&=e^{-i\omega r_*}\sum_{i=0}^{n^H_{\text{max}}}b_i^{(K)}(r-2M_{BH})^i
\end{aligned}
\end{equation}
and we perform a similar expansion at large $r$:
\begin{equation}
\begin{aligned}
\label{eq:inf_BCs}
H_0(r\rightarrow\infty)&=e^{i\omega r_*}\sum_{i=1}^{n^\infty_{\text{max}}}\frac{a_i^{(H_0)}}{r^i},\\
H_1(r\rightarrow\infty)&=e^{i\omega r_*}\sum_{i=1}^{n^\infty_{\text{max}}}\frac{a_i^{(H_1)}}{r^i},\\
K(r\rightarrow\infty)&=e^{i\omega r_*}\sum_{i=0}^{n^\infty_{\text{max}}}\frac{a_i^{(K)}}{r^i}.
\end{aligned}
\end{equation}
The maximum indices in \ref{eq:Hor_BCs} and \ref{eq:inf_BCs} are chosen to be $n^H_{\text{max}}=3$ and $n^\infty_{\text{max}}=5$, sufficient for our desired precision. We assume that the fluid perturbations do not propagate to spatial infinity and that they vanish at the horizon, such that $W^\infty=\delta\rho^\infty=W^H=\delta\rho^H=0$. This assumption is equivalent to the statement that the background density profile is static, i.e., it does not deplete over time due to the interactions between the secondary perturber and the environment. This assumption is sufficient for our purposes, but we refer the reader to Refs.~\cite{Dyson:2025dlj,Li:2025ffh,Brito:2023pyl} for perturbative treatments which incorporate such environmental depletion. 

As discussed in \ref{sec:model_construction}, the metric functions $m(r)$ and $f(r)$ in the asymptotic regions are given by
\begin{equation}
\begin{aligned}
r&\rightarrow2M_{BH}:~~ m= M_{BH},~~~~
f=\left(1-\frac{2M_{BH}}{r}\right)e^{\mu_0},~~~~\\
r&\rightarrow\infty:~~~~~~~m= M_{ADM},~~
f= 1-\frac{2M_{ADM}}{r}.\nonumber
\end{aligned}
\end{equation}
Plugging these expressions into the system of polar equations, we can solve for the coefficients $a_i^{(X)}$ and $b_i^{(X)}$ with $X=\{H_0, H_1, K\}$, up to a normalization of the field $K$. Asymptotically, we normalize $K$ such that $a_0^{(K)}=1$. Doing so, we are left with the amplitude $b_0^{(K)}$ as a free parameter. To compute GW energy fluxes, we employ a shooting method to numerically determine $b_0^{(K)}$ such that the Wronskian $\lim_{r\rightarrow\infty}(K_\text{exp}'K_{\text{num}}-K_\text{exp}K_{\text{num}}')/K_\text{exp}'=0$. Here, $K_\text{exp}$ refers to the expansion in powers of $1/r$, and $K_\text{num}$ refers to the numerically generated solution to the system of polar equations \ref{eq:polar_pert}. Once we have solved for $K$, the GW energy flux at $r \to \infty$ is then computed as
\begin{equation}
\dot{E}_\infty=\frac{1}{32\pi}\frac{(\ell+2)!}{(\ell-2)!}\lim_{r\rightarrow\infty}|K(r)|^2.
\end{equation}
In our numerical treatment, we approximate the spatial infinity result by extracting the fields at a large radius $r_{\text{ext}}=10000/\Omega_p$. This choice ensures that the extraction radius is sufficiently far from the source to encompass many wavelengths of the emitted gravitational radiation, ensuring that $\dot{E}(r_{\rm ext})$ provides a reliable and stable representation for the flux at infinity. 

\begin{figure}[t]
\centering
\includegraphics[width=\linewidth]{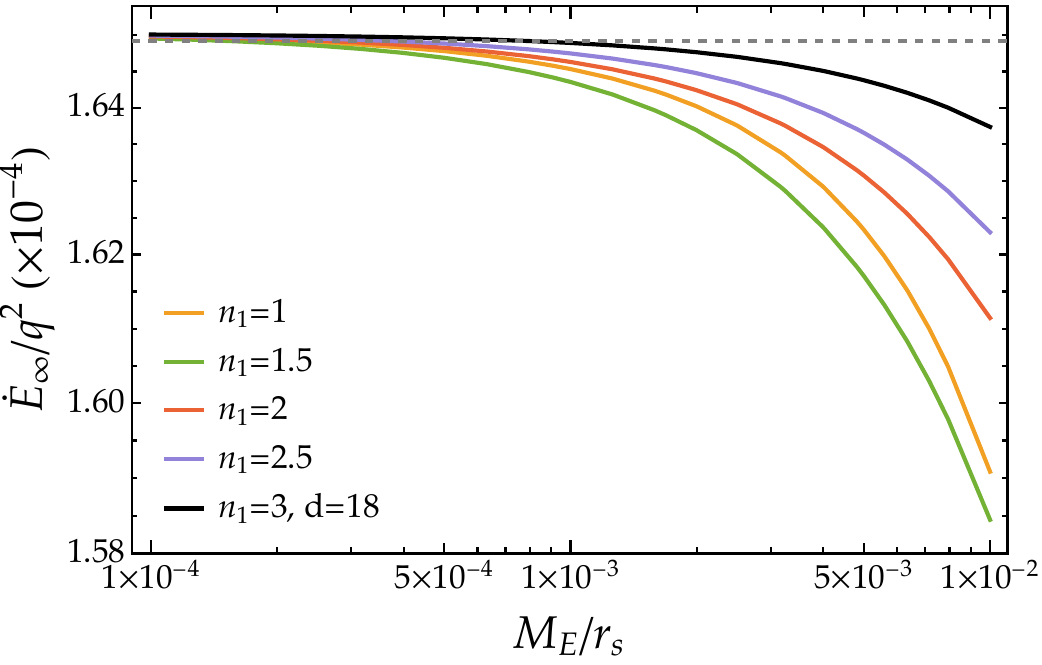}
\caption{Radiated GW energy flux in the $(\ell,m)=(2,2)$ mode at spatial infinity, for a particle in circular orbit at $r_p=8$. Here, we approximate the particle's source with a Gaussian with width $\sigma=1/25$, and fix the fluid sound speeds $(c_{s_t},c_{s_r})=(0,0.9)$. The fluxes are extracted at a radius $r_{\text{ext}}=10000/\Omega_p$, which we choose to be sufficiently outside of the DM halo. The dashed gray line denotes the vacuum Schwarzschild flux for the same circular orbit. The DM halo parameters $\alpha, n_2$, and $n_3$ are the same as in \ref{fig2}.}
\label{fig:GW_flux_vary_compactness}
\end{figure}

\begin{figure}[t]
\centering
\includegraphics[width=\linewidth]{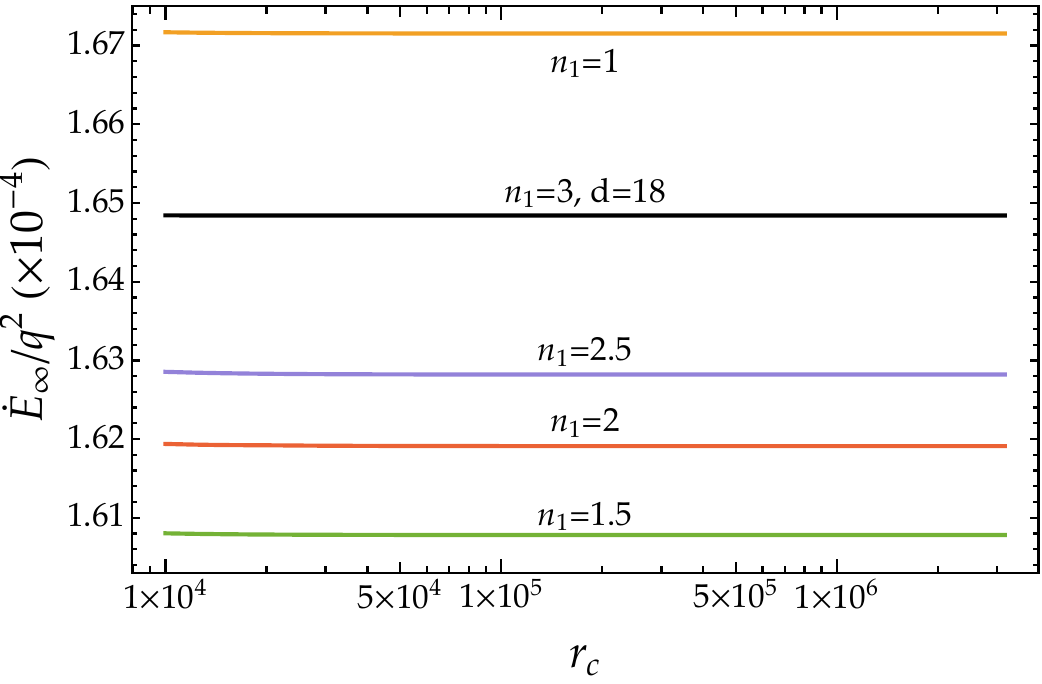}
\caption{Radiated GW energy flux in the $(\ell,m)=(2,2)$ mode at spatial infinity, for a particle in circular orbit at $r_p=8$. All parameters are the same as in \ref{fig:GW_flux_vary_compactness}, except here we fix $M_E=10$, $r_s=10^3$, and vary the profile's cutoff radius $r_c$.}
\label{fig:GW_flux_vary_rc}
\end{figure}

The results of the procedure described above are shown for the $(\ell,m)=(2,2)$ mode in ~\ref{fig:GW_flux_vary_compactness} and ~\ref{fig:GW_flux_vary_rc}, for a particle moving on a circular orbit at $r_p=8$. This radius is chosen since it is sufficiently close to the highest density regions of the DM halo, where fluid couplings are expected to play a more important role. The localization of the point particle source term on the orbit of the secondary body is provided via a Gaussian (rather than a singular delta function) with width $\sigma=1/25$, and the tangential (`t') and radial (`r') sound speeds in the polar perturbation equations are taken to be $(c_{s_t},c_{s_r})=(0,0.9)$ for simplicity. More physically motivated sound speeds would require an exploration of the microphysics of the environment, which is beyond the scope of this work. 

In \ref{fig:GW_flux_vary_compactness}, we fix the scale radius to $r_s=10^4$ and compute the GW flux radiated to infinity as a function of the compactness $z=M_E/r_s$ for various choices of $n_1$ in the parameterized halo model. Small values of $n_1$ for the ${}_2F_1$-model correspond to more steeply-sloped DM density profiles, and to a higher enclosed mass within the particle radius $r_p$. The opposite is true as one increases $n_1$; the halo density profile becomes less steeply-sloped and hence more diffuse around $r_0$ with increasing $n_1$. The GW fluxes shown in \ref{fig:GW_flux_vary_compactness} can be explained through this behavior. 

For all cases, the outgoing GW flux is close to the vacuum value towards the left of the plot, where the compactness is smallest and the halo is most diffused. For high values of $n_1$, as seen from \ref{fig2} as well, the slope of the DM profile is quite shallow, and the density is spread out over large distances. The resulting energy flux at spatial infinity is mostly close to the vacuum value, with very little energy being transferred to the environmental excitations. As the compactness increases, more energy is lost to the fluid perturbations as the inner densities become higher, and hence less energy is radiated away in GWs. This effect becomes more pronounced as $n_1$ decreases, as the inner slope and enclosed mass $m(r_p)$ become higher. The trend eventually breaks when $n_1$ decreases even further, as seen in the $n_1=1$ case. Now the particle ``feels'' an enclosed mass that is a significant fraction of the central BH mass ($m_E(r_p)\sim0.02M_{BH}$ at $z=10^{-2}$ for these parameters), and hence the change in the particle's binding energy and orbital frequency becomes significant, leading to an increase in the GW flux. 

While realistic DM halos are closer to the left-hand side of the parameter space explored in \ref{fig:GW_flux_vary_compactness}, it is apparent that low values of $n_1$ will stand out more in future EMRI GW measurements, as they lead to considerable changes in the emitted GW flux from the vacuum counterpart, especially for higher compactness. However, without a more comprehensive parameter estimation study of non-vacuum EMRI signals in our parameterized framework, we cannot fully determine the viability of such parameter constraints through the behavior of the fluxes alone. Indeed, additional complications and possible degeneracies are expected to arise when considering spinning primary masses and eccentric orbits. 

In \ref{fig:GW_flux_vary_rc} we show how the GW fluxes behave as a function of the cutoff radius $r_c$. Unsurprisingly, the radiated GW energy flux has an extremely mild dependence on $r_c$, because the extraction radius $r_{\text{ext}}$ is always larger than the halo cutoff. There is a mild variation in the fluxes for low values of $r_c$ when $r_{c}$ is small, and hence very close to the scale radius $r_s$. Also, as discussed earlier, increasing $n_1$ produces more ``evenly spread” halos, resulting in weaker excitation of the fluid modes and, consequently, in a larger fraction of the energy being radiated to infinity. This trend, however, reverses as $n_1$ decreases below a certain limit. In this regime, the enclosed mass $m(r_p)$ and the corresponding orbital frequency $\Omega_p$ increase sufficiently rapidly that the enhancement in flux overcomes the suppression due to energy deposited into the fluid modes, leading to an overall increase in $\dot{E}_\infty$. 

\section{Conclusions}
In this work, we develop a general framework for describing BHs embedded in DM halos by parametrizing the spacetime directly in terms of the enclosed mass profile, rather than the underlying density distribution. Since gravitational observables are sensitive primarily to the cumulative mass distribution, this formulation provides a natural and observationally motivated characterization of the environment, while remaining agnostic about its fine-grained details. The framework also incorporates key physical consistency requirements, including regularity, positivity of density, causality, and appropriate asymptotic behavior. At the same time, it encompasses a broad class of commonly employed density profiles as limiting cases, allowing a diverse range of astrophysical environments to be described within a unified framework.

We have applied the model to investigate environmental effects on key GW observables, including QNMs, TLNs, and GW fluxes from EMRIs. Our results indicate that different GW observables probe the environmental mass distribution in qualitatively different ways. The QNM spectrum is primarily governed by the inner-halo profile and the light ring region, whereas static TLNs encode information accumulated over a much broader radial domain, and therefore they are sensitive to the outer structure of the halo. The EMRI fluxes provide a complementary dynamical probe, as the orbital motion responds directly to the enclosed mass at the particle’s location, while the subsequent redshifted-propagation of perturbations and excitation of the fluid degrees of freedom determine the fraction of the emitted radiation reaching infinity. Taken together, these observables provide complementary information about the radial structure of the environment, suggesting that their combination could help break degeneracies among halo parameters that may remain unresolved when considering any single observable in isolation. 

As GW detectors continue to improve in sensitivity, and in particular with the advent of next-generation ground-based observatories and space-based missions targeting EMRIs, these observables may offer valuable insights into the structure of DM halos and the interplay among the model parameters. Ultimately, this could enable tighter constraints on the halo properties and improve our understanding of the astrophysical effects of matter environments on BH dynamics and GW observables.

This work naturally opens several directions for future investigation. Although our analysis has been restricted to static, spherically symmetric configurations, the formalism is sufficiently general to accommodate additional physical effects, such as rotation and multi-component environments. It would also be interesting to extend the analysis to dynamical TLNs, incorporating frequency-dependent effects through an expansion in $(\mathcal{R}\,\omega)$, where $\mathcal{R}$ denotes an appropriate characteristic length scale of the system. More broadly, this provides a natural setting for developing an effective field theory description of tidal response in the presence of an extended environment.

On the observational side, confronting the model with GW data through parameter estimation and Bayesian inference will be essential for assessing its measurability and constraining the properties of the surrounding matter. More broadly, the enclosed-mass approach developed here provides a versatile bridge between astrophysical modeling and precision tests of strong-field gravity, offering a unified framework for exploring the phenomenology of BHs in realistic astrophysical environments.

\section*{Acknowledgements}
We thank Andrea Maselli for valuable feedback on the draft. 
R.G. is supported by the Fulbright Nehru Postdoctoral Research Fellowship (Award No.3174/FNPDR/2025) from the United States-India Educational Foundation.
R.G. and E.B. are supported by NSF Grants No.~AST-2606672, No.~PHY-2513337, No.~PHY-090003, and No.~PHY-20043, by John Templeton Foundation Grant No.~62840, by the Simons Foundation [MPS-SIP-00001698, E.B.], by the Simons Foundation International [SFI-MPS-BH-00012593-02], and by Italian Ministry of Foreign Affairs and International Cooperation Grant No.~PGR01167. 
Part of this work was carried out at the Advanced Research Computing at Hopkins (ARCH) core facility (\url{https://www.arch.jhu.edu/}), which is supported by the NSF Grant No. OAC-1920103.
R.G., A.C. and C.S. acknowledge the hospitality of IIT Gandhinagar and the stimulating discussions during the program ``GW10@IITGN,’’ which played an important role in the initial phase of this work. MASP acknowledges funding from the Funda\c{c}\~{a}o para a Ci\^{e}ncia e a Tecnologia (FCT) grant UID/04434/2025 and through the Fellowship UI/BD/154479/2022. 
MASP also acknowledges financial support from the Luso-American Development Foundation (FLAD) under an R\&D Internship Grant, from a Fulbright Grant for Research, supported by FCT, and support from the Spanish Grant PID2023-149560NB-C21, funded by MCIN/AEI/10.13039/501100011033.

\bibliography{reference}

\providecommand{\href}[2]{#2}\begingroup\raggedright\begin{thebibliography}{10}

\bibitem{j_binney_galactic_1987}
{J. Binney} and {S. Tremaine}, {\em Galactic dynamics}.
\newblock Princeton series in astrophysics. Princeton University Press, Princeton, N.J., 1987.

\bibitem{1996Morris}
M.~{Morris} and E.~{Serabyn}, ``{The Galactic Center Environment},'' \href{http://dx.doi.org/10.1146/annurev.astro.34.1.645}{{\em Annual Review of Astronomy and Astrophysics} {\bfseries 34} (Jan., 1996) 645--702}.

\bibitem{mo2010galaxy}
H.~Mo, F.~C. van~den Bosch, and S.~White, \href{http://dx.doi.org/10.1017/CBO9780511807244}{{\em Galaxy Formation and Evolution}}.
\newblock Cambridge University Press, Cambridge, UK, 2010.

\bibitem{Feng:2010gw}
J.~L. Feng, ``{Dark Matter Candidates from Particle Physics and Methods of Detection},'' \href{http://dx.doi.org/10.1146/annurev-astro-082708-101659}{{\em Annu. Rev. Astron. Astrophys.} {\bfseries 48} (2010) 495--545}, \href{http://arxiv.org/abs/1003.0904}{{\ttfamily arXiv:1003.0904 [astro-ph.CO]}}.

\bibitem{Alexander:2016aln}
J.~Alexander {\em et~al.}, ``{Dark Sectors 2016 Workshop: Community Report},''
\newblock 8, 2016.
\newblock \href{http://arxiv.org/abs/1608.08632}{{\ttfamily arXiv:1608.08632 [hep-ph]}}.

\bibitem{Ferreira:2020fam}
E.~G.~M. Ferreira, ``{Ultra-light dark matter},'' \href{http://dx.doi.org/10.1007/s00159-021-00135-6}{{\em Astron. Astrophys. Rev.} {\bfseries 29} no.~1, (2021) 7}, \href{http://arxiv.org/abs/2005.03254}{{\ttfamily arXiv:2005.03254 [astro-ph.CO]}}.

\bibitem{Barausse:2014tra}
E.~Barausse, V.~Cardoso, and P.~Pani, ``{Can environmental effects spoil precision gravitational-wave astrophysics?},'' \href{http://dx.doi.org/10.1103/PhysRevD.89.104059}{{\em Phys. Rev. D} {\bfseries 89} no.~10, (2014) 104059}, \href{http://arxiv.org/abs/1404.7149}{{\ttfamily arXiv:1404.7149 [gr-qc]}}.

\bibitem{Cardoso:2021wlq}
V.~Cardoso, K.~Destounis, F.~Duque, R.~P. Macedo, and A.~Maselli, ``{Black holes in galaxies: Environmental impact on gravitational-wave generation and propagation},'' \href{http://dx.doi.org/10.1103/PhysRevD.105.L061501}{{\em Phys. Rev. D} {\bfseries 105} no.~6, (2022) L061501}, \href{http://arxiv.org/abs/2109.00005}{{\ttfamily arXiv:2109.00005 [gr-qc]}}.

\bibitem{Speeney:2022ryg}
N.~Speeney, A.~Antonelli, V.~Baibhav, and E.~Berti, ``{Impact of relativistic corrections on the detectability of dark-matter spikes with gravitational waves},'' \href{http://dx.doi.org/10.1103/PhysRevD.106.044027}{{\em Phys. Rev. D} {\bfseries 106} no.~4, (2022) 044027}, \href{http://arxiv.org/abs/2204.12508}{{\ttfamily arXiv:2204.12508 [gr-qc]}}.

\bibitem{Zhao:2023tyo}
Y.~Zhao, B.~Sun, K.~Lin, and Z.~Cao, ``{Axial gravitational ringing of a spherically symmetric black hole surrounded by dark matter spike},'' \href{http://dx.doi.org/10.1103/PhysRevD.108.024070}{{\em Phys. Rev. D} {\bfseries 108} no.~2, (2023) 024070}, \href{http://arxiv.org/abs/2303.09215}{{\ttfamily arXiv:2303.09215 [gr-qc]}}.

\bibitem{Speeney:2024mas}
N.~Speeney, E.~Berti, V.~Cardoso, and A.~Maselli, ``{Black holes surrounded by generic matter distributions: Polar perturbations and energy flux},'' \href{http://dx.doi.org/10.1103/PhysRevD.109.084068}{{\em Phys. Rev. D} {\bfseries 109} no.~8, (2024) 084068}, \href{http://arxiv.org/abs/2401.00932}{{\ttfamily arXiv:2401.00932 [gr-qc]}}.

\bibitem{Chakraborty:2024gcr}
S.~Chakraborty, G.~Comp{\`e}re, and L.~Machet, ``{Tidal Love numbers and quasinormal modes of the Schwarzschild-Hernquist black hole},'' \href{http://dx.doi.org/10.1103/4p2c-rwdh}{{\em Phys. Rev. D} {\bfseries 112} no.~2, (2025) 024015}, \href{http://arxiv.org/abs/2412.14831}{{\ttfamily arXiv:2412.14831 [gr-qc]}}.

\bibitem{DOnofrio:2026ulh}
S.~D'Onofrio, S.~Datta, and A.~Maselli, ``{Axial tidal Love numbers of black holes in matter environments},'' \href{http://arxiv.org/abs/2605.02633}{{\ttfamily arXiv:2605.02633 [gr-qc]}}.

\bibitem{Chowdhury:2025tpt}
A.~Chowdhury, G.~Sen, S.~Chakrabarti, and S.~Das, ``{Effect of generic dark matter halos on the transonic accretion onto galactic black holes},'' \href{http://dx.doi.org/10.1103/vqjm-3dt8}{{\em Phys. Rev. D} {\bfseries 112} no.~6, (2025) 064041}, \href{http://arxiv.org/abs/2503.08528}{{\ttfamily arXiv:2503.08528 [gr-qc]}}.

\bibitem{Pezzella:2024tkf}
L.~Pezzella, K.~Destounis, A.~Maselli, and V.~Cardoso, ``{Quasinormal modes of black holes embedded in halos of matter},'' \href{http://dx.doi.org/10.1103/PhysRevD.111.064026}{{\em Phys. Rev. D} {\bfseries 111} no.~6, (2025) 064026}, \href{http://arxiv.org/abs/2412.18651}{{\ttfamily arXiv:2412.18651 [gr-qc]}}.

\bibitem{Kavanagh:2020cfn}
B.~J. Kavanagh, D.~A. Nichols, G.~Bertone, and D.~Gaggero, ``{Detecting dark matter around black holes with gravitational waves: Effects of dark-matter dynamics on the gravitational waveform},'' \href{http://dx.doi.org/10.1103/PhysRevD.102.083006}{{\em Phys. Rev. D} {\bfseries 102} no.~8, (2020) 083006}, \href{http://arxiv.org/abs/2002.12811}{{\ttfamily arXiv:2002.12811 [gr-qc]}}.

\bibitem{Duque:2023seg}
F.~Duque, C.~F.~B. Macedo, R.~Vicente, and V.~Cardoso, ``{Extreme-Mass-Ratio Inspirals in Ultralight Dark Matter},'' \href{http://dx.doi.org/10.1103/PhysRevLett.133.121404}{{\em Phys. Rev. Lett.} {\bfseries 133} no.~12, (2024) 121404}, \href{http://arxiv.org/abs/2312.06767}{{\ttfamily arXiv:2312.06767 [gr-qc]}}.

\bibitem{Barausse:2014pra}
E.~Barausse, V.~Cardoso, and P.~Pani, ``{Environmental Effects for Gravitational-wave Astrophysics},'' \href{http://dx.doi.org/10.1088/1742-6596/610/1/012044}{{\em J. Phys. Conf. Ser.} {\bfseries 610} no.~1, (2015) 012044}, \href{http://arxiv.org/abs/1404.7140}{{\ttfamily arXiv:1404.7140 [astro-ph.CO]}}.

\bibitem{Zhang:2024ugv}
C.~Zhang, G.~Fu, and N.~Dai, ``{Detecting dark matter halos with extreme mass-ratio inspirals},'' \href{http://dx.doi.org/10.1088/1475-7516/2024/04/088}{{\em J. Cosmology Astropart. Phys.} {\bfseries 04} (2024) 088}, \href{http://arxiv.org/abs/2401.04467}{{\ttfamily arXiv:2401.04467 [gr-qc]}}.

\bibitem{Gliorio:2025cbh}
S.~Gliorio, E.~Berti, A.~Maselli, and N.~Speeney, ``{Extreme mass ratio inspirals in dark matter halos: Dynamics and distinguishability of halo models},'' \href{http://dx.doi.org/10.1103/dw6c-14pt}{{\em Phys. Rev. D} {\bfseries 112} no.~12, (2025) 124050}, \href{http://arxiv.org/abs/2503.16649}{{\ttfamily arXiv:2503.16649 [gr-qc]}}.

\bibitem{Mitra:2025tag}
S.~Mitra, N.~Speeney, S.~Chakraborty, and E.~Berti, ``{Extreme mass ratio inspirals in rotating dark matter spikes},'' \href{http://dx.doi.org/10.1103/ycl1-kx7d}{{\em Phys. Rev. D} {\bfseries 112} no.~4, (2025) 044030}, \href{http://arxiv.org/abs/2505.04697}{{\ttfamily arXiv:2505.04697 [gr-qc]}}.

\bibitem{Fonseca:2025ehf}
D.~S. Fonseca, C.~F.~B. Macedo, M.~Malato~Corr{\^e}a, and D.~Rubiera-Garcia, ``{Matter environments around black holes: Geodesics, light rings, and ultracompact configurations},'' \href{http://dx.doi.org/10.1103/x95d-ry14}{{\em Phys. Rev. D} {\bfseries 113} no.~12, (2026) 124039}, \href{http://arxiv.org/abs/2512.22267}{{\ttfamily arXiv:2512.22267 [gr-qc]}}.

\bibitem{Macedo:2024qky}
C.~F.~B. Macedo, J.~a.~L. Rosa, and D.~Rubiera-Garcia, ``{Optical appearance of black holes surrounded by a dark matter halo},'' \href{http://dx.doi.org/10.1088/1475-7516/2024/07/046}{{\em J. Cosmology Astropart. Phys.} {\bfseries 07} (2024) 046}, \href{http://arxiv.org/abs/2402.13047}{{\ttfamily arXiv:2402.13047 [gr-qc]}}.

\bibitem{Bhowmik:2026owi}
A.~Bhowmik, A.~Chowdhury, and S.~Chakrabarti, ``{The Ringdown and the Tide: Fingerprints of Dark Matter Halo Profiles},'' \href{http://arxiv.org/abs/2608.07678}{{\ttfamily arXiv:2608.07678 [gr-qc]}}.

\bibitem{Cardoso:2019rvt}
V.~Cardoso and P.~Pani, ``{Testing the nature of dark compact objects: a status report},'' \href{http://dx.doi.org/10.1007/s41114-019-0020-4}{{\em Living Rev. Relativ.} {\bfseries 22} no.~1, (2019) 4}, \href{http://arxiv.org/abs/1904.05363}{{\ttfamily arXiv:1904.05363 [gr-qc]}}.

\bibitem{Amaro-Seoane:2012lgq}
P.~Amaro-Seoane, ``{Relativistic dynamics and extreme mass ratio inspirals},'' \href{http://dx.doi.org/10.1007/s41114-018-0013-8}{{\em Living Rev. Relativ.} {\bfseries 21} no.~1, (2018) 4}, \href{http://arxiv.org/abs/1205.5240}{{\ttfamily arXiv:1205.5240 [astro-ph.CO]}}.

\bibitem{Babak:2017tow}
S.~Babak, J.~Gair, A.~Sesana, E.~Barausse, C.~F. Sopuerta, C.~P.~L. Berry, E.~Berti, P.~Amaro-Seoane, A.~Petiteau, and A.~Klein, ``{Science with the space-based interferometer LISA. V: Extreme mass-ratio inspirals},'' \href{http://dx.doi.org/10.1103/PhysRevD.95.103012}{{\em Phys. Rev. D} {\bfseries 95} no.~10, (2017) 103012}, \href{http://arxiv.org/abs/1703.09722}{{\ttfamily arXiv:1703.09722 [gr-qc]}}.

\bibitem{Navarro:1996gj}
J.~F. Navarro, C.~S. Frenk, and S.~D.~M. White, ``{A Universal density profile from hierarchical clustering},'' \href{http://dx.doi.org/10.1086/304888}{{\em Astrophys. J.} {\bfseries 490} (1997) 493--508}, \href{http://arxiv.org/abs/astro-ph/9611107}{{\ttfamily arXiv:astro-ph/9611107}}.

\bibitem{1969Afz.....5..137E}
J.~{Einasto}, ``{The Andromeda galaxy M31. I. A preliminary model},'' {\em Astrofizika} {\bfseries 5} (Feb., 1969) 137--159.

\bibitem{Hernquist:1990be}
L.~Hernquist, ``{An Analytical Model for Spherical Galaxies and Bulges},'' \href{http://dx.doi.org/10.1086/168845}{{\em Astrophys. J.} {\bfseries 356} (1990) 359}.

\bibitem{Burkert:1995yz}
A.~Burkert, ``{The Structure of dark matter halos in dwarf galaxies},'' \href{http://dx.doi.org/10.1086/309560}{{\em Astrophys. J. Lett.} {\bfseries 447} (1995) L25}, \href{http://arxiv.org/abs/astro-ph/9504041}{{\ttfamily arXiv:astro-ph/9504041}}.

\bibitem{Gondolo:1999ef}
P.~Gondolo and J.~Silk, ``{Dark matter annihilation at the galactic center},'' \href{http://dx.doi.org/10.1103/PhysRevLett.83.1719}{{\em Phys. Rev. Lett.} {\bfseries 83} (1999) 1719--1722}, \href{http://arxiv.org/abs/astro-ph/9906391}{{\ttfamily arXiv:astro-ph/9906391}}.

\bibitem{Sadeghian:2013laa}
L.~Sadeghian, F.~Ferrer, and C.~M. Will, ``{Dark matter distributions around massive black holes: A general relativistic analysis},'' \href{http://dx.doi.org/10.1103/PhysRevD.88.063522}{{\em Phys. Rev. D} {\bfseries 88} no.~6, (2013) 063522}, \href{http://arxiv.org/abs/1305.2619}{{\ttfamily arXiv:1305.2619 [astro-ph.GA]}}.

\bibitem{Ferrer:2017xwm}
F.~Ferrer, A.~M. da~Rosa, and C.~M. Will, ``{Dark matter spikes in the vicinity of Kerr black holes},'' \href{http://dx.doi.org/10.1103/PhysRevD.96.083014}{{\em Phys. Rev. D} {\bfseries 96} no.~8, (2017) 083014}, \href{http://arxiv.org/abs/1707.06302}{{\ttfamily arXiv:1707.06302 [astro-ph.CO]}}.

\bibitem{Poisson:2009pwt}
E.~Poisson, \href{http://dx.doi.org/10.1017/CBO9780511606601}{{\em {A Relativist's Toolkit: The Mathematics of Black-Hole Mechanics}}}.
\newblock Cambridge University Press, 12, 2009.

\bibitem{Persic:1995ru}
M.~Persic, P.~Salucci, and F.~Stel, ``{The Universal rotation curve of spiral galaxies: 1. The Dark matter connection},'' \href{http://dx.doi.org/10.1093/mnras/278.1.27}{{\em Mon. Not. R. Astron. Soc.} {\bfseries 281} (1996) 27}, \href{http://arxiv.org/abs/astro-ph/9506004}{{\ttfamily arXiv:astro-ph/9506004}}.

\bibitem{Bertone:2016nfn}
G.~Bertone and D.~Hooper, ``{History of dark matter},'' \href{http://dx.doi.org/10.1103/RevModPhys.90.045002}{{\em Rev. Mod. Phys.} {\bfseries 90} no.~4, (2018) 045002}, \href{http://arxiv.org/abs/1605.04909}{{\ttfamily arXiv:1605.04909 [astro-ph.CO]}}.

\bibitem{2010MNRAS.406.1220W}
J.~{Wolf}, G.~D. {Martinez}, J.~S. {Bullock}, M.~{Kaplinghat}, M.~{Geha}, R.~R. {Mu{\~n}oz}, J.~D. {Simon}, and F.~F. {Avedo}, ``{Accurate masses for dispersion-supported galaxies},'' \href{http://dx.doi.org/10.1111/j.1365-2966.2010.16753.x}{{\em Mon. Not. R. Astron. Soc.} {\bfseries 406} no.~2, (Aug., 2010) 1220--1237}, \href{http://arxiv.org/abs/0908.2995}{{\ttfamily arXiv:0908.2995 [astro-ph.CO]}}.

\bibitem{1992grle.book.....S}
P.~{Schneider}, J.~{Ehlers}, and E.~E. {Falco}, \href{http://dx.doi.org/10.1007/978-3-662-03758-4}{{\em {Gravitational Lenses}}}.
\newblock Springer, 1992.

\bibitem{Cardoso:2019upw}
V.~Cardoso and F.~Duque, ``{Environmental effects in gravitational-wave physics: Tidal deformability of black holes immersed in matter},'' \href{http://dx.doi.org/10.1103/PhysRevD.101.064028}{{\em Phys. Rev. D} {\bfseries 101} no.~6, (2020) 064028}, \href{http://arxiv.org/abs/1912.07616}{{\ttfamily arXiv:1912.07616 [gr-qc]}}.

\bibitem{Chakravarti:2025awj}
K.~Chakravarti and C.~Singha, ``{Tidal Love numbers and quasi-normal modes of the ECO in a Dark Matter halo},'' \href{http://arxiv.org/abs/2509.03556}{{\ttfamily arXiv:2509.03556 [gr-qc]}}.

\bibitem{Zhao:2026eti}
Y.-Q. Zhao and P.~Pani, ``{Quasinormal modes and tidal responses of black holes in generic anisotropic matter environments},'' \href{http://arxiv.org/abs/2606.11380}{{\ttfamily arXiv:2606.11380 [gr-qc]}}.

\bibitem{Chakraborty:2026qru}
S.~Chakraborty and P.~Pani, ``{Tidal Response of Compact Objects},'' \href{http://arxiv.org/abs/2604.08679}{{\ttfamily arXiv:2604.08679 [gr-qc]}}.

\bibitem{Cannizzaro:2024fpz}
E.~Cannizzaro, V.~De~Luca, and P.~Pani, ``{Tidal deformability of black holes surrounded by thin accretion disks},'' \href{http://dx.doi.org/10.1103/PhysRevD.110.123004}{{\em Phys. Rev. D} {\bfseries 110} no.~12, (2024) 123004}, \href{http://arxiv.org/abs/2408.14208}{{\ttfamily arXiv:2408.14208 [astro-ph.HE]}}.

\bibitem{Chowdhury:2026cjv}
A.~Chowdhury, C.~Singha, K.~Bamba, and S.~Chakraborty, ``{Tidal deformation of an accreting compact object},'' \href{http://arxiv.org/abs/2607.24938}{{\ttfamily arXiv:2607.24938 [gr-qc]}}.

\bibitem{jebsen}
J.~Jebsen, {\em {\"U}ber die allgemeinen kugelsymmetrischen L{\"o}sungen der Einstein'schen Gravitationsgleichungen im Vakuum}.
\newblock Arkiv f{\"o}r Matematik, Astronomi och Fysik. Almqvist \& Wiksell, 1921.

\bibitem{birkhoff}
G.~D. {Birkhoff} and R.~E. {Langer}, {\em {Relativity and modern physics}}.
\newblock Harvard University Press, 1923.

\bibitem{Misner:1964je}
C.~W. Misner and D.~H. Sharp, ``{Relativistic Equations for Adiabatic, Spherically Symmetric Gravitational Collapse},'' \href{http://dx.doi.org/10.1103/PhysRev.136.B571}{{\em Phys. Rev.} {\bfseries 136} (1964) B571--B576}.

\bibitem{Hayward:1994bu}
S.~A. Hayward, ``{Gravitational energy in spherical symmetry},'' \href{http://dx.doi.org/10.1103/PhysRevD.53.1938}{{\em Phys. Rev. D} {\bfseries 53} (1996) 1938--1949}, \href{http://arxiv.org/abs/gr-qc/9408002}{{\ttfamily arXiv:gr-qc/9408002}}.

\bibitem{Vishveshwara1968}
C.~V. Vishveshwara, ``Generalization of the ``schwarzschild surface'' to arbitrary static and stationary metrics,'' \href{http://dx.doi.org/10.1063/1.1664723}{{\em Journal of Mathematical Physics} {\bfseries 9} no.~8, (1968) 1319--1329}. \url{https://pubs.aip.org/aip/jmp/article/9/8/1319/234397/Generalization-of-the-Schwarzschild-Surface-to}.

\bibitem{MSM_1927__25__1_0}
G.~Darmois, {\em The Equations of Einsteinian Gravitation}.
\newblock No.~25 in M\'emorial des sciences math\'ematiques. Gauthier-Villars, 1927.

\bibitem{Israel:1966rt}
W.~Israel, ``{Singular hypersurfaces and thin shells in general relativity},'' \href{http://dx.doi.org/10.1007/BF02710419}{{\em Nuovo Cim. B} {\bfseries 44S10} (1966) 1}. [Erratum: Nuovo Cim.B 48, 463 (1967)].

\bibitem{Wald:1984rg}
R.~M. Wald, \href{http://dx.doi.org/10.7208/chicago/9780226870373.001.0001}{{\em {General Relativity}}}.
\newblock Chicago Univ. Pr., Chicago, USA, 1984.

\bibitem{Bardeen:1973gs}
J.~M. Bardeen, B.~Carter, and S.~W. Hawking, ``{The Four laws of black hole mechanics},'' \href{http://dx.doi.org/10.1007/BF01645742}{{\em Commun. Math. Phys.} {\bfseries 31} (1973) 161--170}.

\bibitem{Alho:2021sli}
A.~Alho, J.~Nat{\'a}rio, P.~Pani, and G.~Raposo, ``{Compact elastic objects in general relativity},'' \href{http://dx.doi.org/10.1103/PhysRevD.105.044025}{{\em Phys. Rev. D} {\bfseries 105} no.~4, (2022) 044025}, \href{http://arxiv.org/abs/2107.12272}{{\ttfamily arXiv:2107.12272 [gr-qc]}}. [Erratum: Phys.Rev.D 105, 129903 (2022)].

\bibitem{Datta:2023zmd}
S.~Datta, ``{Black holes immersed in dark matter: Energy condition and sound speed},'' \href{http://dx.doi.org/10.1103/PhysRevD.109.104042}{{\em Phys. Rev. D} {\bfseries 109} no.~10, (2024) 104042}, \href{http://arxiv.org/abs/2312.01277}{{\ttfamily arXiv:2312.01277 [gr-qc]}}.

\bibitem{Pani:2025qxs}
P.~Pani, M.~M. Riva, L.~Santoni, N.~Savi{\'c}, and F.~Vernizzi, ``{Nonlinear relativistic tidal response of neutron stars},'' \href{http://dx.doi.org/10.1007/JHEP05(2026)074}{{\em J. High Energy Phys.} {\bfseries 05} (2026) 074}, \href{http://arxiv.org/abs/2512.14663}{{\ttfamily arXiv:2512.14663 [gr-qc]}}.

\bibitem{abramowitz1964handbook}
M.~Abramowitz and I.~Stegun, {\em Handbook of Mathematical Functions: With Formulas, Graphs, and Mathematical Tables}.
\newblock Applied mathematics series. Dover Publications, 1965.

\bibitem{NIST:DLMF}
``{\it NIST Digital Library of Mathematical Functions}.''
\newblock \url{https://dlmf.nist.gov/}. F.~W.~J. Olver, A.~B. {Olde Daalhuis}, D.~W. Lozier, B.~I. Schneider, R.~F. Boisvert, C.~W. Clark, B.~R. Miller, B.~V. Saunders, H.~S. Cohl, and M.~A. McClain, eds.

\bibitem{Arfken2012MathematicalMF}
G.~B. Arfken, H.~J. Weber, and F.~E. Harris, {\em Mathematical Methods for Physicists: A Comprehensive Guide}.
\newblock Academic Press, Oxford, 7~ed., 2012.

\bibitem{book:91229966}
R.~R. George E.~Andrews, Richard~Askey, {\em Special Functions}.
\newblock Encyclopedia of Mathematics and its Applications. Cambridge University Press, 1999.

\bibitem{Taylor:2002zd}
J.~E. Taylor and J.~Silk, ``{The Clumpiness of cold dark matter: Implications for the annihilation signal},'' \href{http://dx.doi.org/10.1046/j.1365-8711.2003.06201.x}{{\em Mon. Not. R. Astron. Soc.} {\bfseries 339} (2003) 505}, \href{http://arxiv.org/abs/astro-ph/0207299}{{\ttfamily arXiv:astro-ph/0207299}}.

\bibitem{2020MNRAS.499.2912F}
J.~{Freundlich}, F.~{Jiang}, A.~{Dekel}, N.~{Cornuault}, O.~{Ginzburg}, R.~{Koskas}, S.~{Lapiner}, A.~{Dutton}, and A.~V. {Macci{\`o}}, ``{The Dekel-Zhao profile: a mass-dependent dark-matter density profile with flexible inner slope and analytic potential, velocity dispersion, and lensing properties},'' \href{http://dx.doi.org/10.1093/mnras/staa2790}{{\em Mon. Not. R. Astron. Soc.} {\bfseries 499} no.~2, (Dec., 2020) 2912--2933}, \href{http://arxiv.org/abs/2004.08395}{{\ttfamily arXiv:2004.08395 [astro-ph.GA]}}.

\bibitem{Sersic1963}
J.~L. S{\'e}rsic, ``Influence of the atmospheric and instrumental dispersion on the brightness distribution in a galaxy,'' {\em Boletin de la Asociacion Argentina de Astronomia} {\bfseries 6} (1963) 41--43.

\bibitem{Ghosh:2021txu}
R.~Ghosh and S.~Sarkar, ``{Light rings of stationary spacetimes},'' \href{http://dx.doi.org/10.1103/PhysRevD.104.044019}{{\em Phys. Rev. D} {\bfseries 104} no.~4, (2021) 044019}, \href{http://arxiv.org/abs/2107.07370}{{\ttfamily arXiv:2107.07370 [gr-qc]}}.

\bibitem{Ghosh:2023kge}
R.~Ghosh, S.~Sk, and S.~Sarkar, ``{Hairy black holes: Nonexistence of short hairs and a bound on the light ring size},'' \href{http://dx.doi.org/10.1103/PhysRevD.108.L041501}{{\em Phys. Rev. D} {\bfseries 108} no.~4, (2023) L041501}, \href{http://arxiv.org/abs/2306.14193}{{\ttfamily arXiv:2306.14193 [gr-qc]}}.

\bibitem{Hinderer:2007mb}
T.~Hinderer, ``{Tidal Love numbers of neutron stars},'' \href{http://dx.doi.org/10.1086/533487}{{\em Astrophys. J.} {\bfseries 677} (2008) 1216--1220}, \href{http://arxiv.org/abs/0711.2420}{{\ttfamily arXiv:0711.2420 [astro-ph]}}. [Erratum: Astrophys.J. 697, 964 (2009)].

\bibitem{Binnington:2009bb}
T.~Binnington and E.~Poisson, ``{Relativistic theory of tidal Love numbers},'' \href{http://dx.doi.org/10.1103/PhysRevD.80.084018}{{\em Phys. Rev. D} {\bfseries 80} (2009) 084018}, \href{http://arxiv.org/abs/0906.1366}{{\ttfamily arXiv:0906.1366 [gr-qc]}}.

\bibitem{Damour:2009vw}
T.~Damour and A.~Nagar, ``{Relativistic tidal properties of neutron stars},'' \href{http://dx.doi.org/10.1103/PhysRevD.80.084035}{{\em Phys. Rev. D} {\bfseries 80} (2009) 084035}, \href{http://arxiv.org/abs/0906.0096}{{\ttfamily arXiv:0906.0096 [gr-qc]}}.

\bibitem{Kol:2011vg}
B.~Kol and M.~Smolkin, ``{Black hole stereotyping: Induced gravito-static polarization},'' \href{http://dx.doi.org/10.1007/JHEP02(2012)010}{{\em J. High Energy Phys.} {\bfseries 02} (2012) 010}, \href{http://arxiv.org/abs/1110.3764}{{\ttfamily arXiv:1110.3764 [hep-th]}}.

\bibitem{2013arXiv1304.2228C}
S.~Chakrabarti, T.~Delsate, and J.~Steinhoff, ``{New perspectives on neutron star and black hole spectroscopy and dynamic tides},'' \href{http://arxiv.org/abs/1304.2228}{{\ttfamily arXiv:1304.2228 [gr-qc]}}.

\bibitem{Gurlebeck:2015xpa}
N.~G\"urlebeck, ``{No-hair theorem for Black Holes in Astrophysical Environments},'' \href{http://dx.doi.org/10.1103/PhysRevLett.114.151102}{{\em Phys. Rev. Lett.} {\bfseries 114} no.~15, (2015) 151102}, \href{http://arxiv.org/abs/1503.03240}{{\ttfamily arXiv:1503.03240 [gr-qc]}}.

\bibitem{Chia:2020yla}
H.~S. Chia, ``{Tidal deformation and dissipation of rotating black holes},'' \href{http://dx.doi.org/10.1103/PhysRevD.104.024013}{{\em Phys. Rev. D} {\bfseries 104} no.~2, (2021) 024013}, \href{http://arxiv.org/abs/2010.07300}{{\ttfamily arXiv:2010.07300 [gr-qc]}}.

\bibitem{Bhatt:2023zsy}
R.~P. Bhatt, S.~Chakraborty, and S.~Bose, ``{Addressing issues in defining the Love numbers for black holes},'' \href{http://dx.doi.org/10.1103/PhysRevD.108.084013}{{\em Phys. Rev. D} {\bfseries 108} no.~8, (2023) 084013}, \href{http://arxiv.org/abs/2306.13627}{{\ttfamily arXiv:2306.13627 [gr-qc]}}.

\bibitem{LeTiec:2020bos}
A.~Le~Tiec, M.~Casals, and E.~Franzin, ``{Tidal Love Numbers of Kerr Black Holes},'' \href{http://dx.doi.org/10.1103/PhysRevD.103.084021}{{\em Phys. Rev. D} {\bfseries 103} no.~8, (2021) 084021}, \href{http://arxiv.org/abs/2010.15795}{{\ttfamily arXiv:2010.15795 [gr-qc]}}.

\bibitem{Charalambous:2021mea}
P.~Charalambous, S.~Dubovsky, and M.~M. Ivanov, ``{On the Vanishing of Love Numbers for Kerr Black Holes},'' \href{http://dx.doi.org/10.1007/JHEP05(2021)038}{{\em J. High Energy Phys.} {\bfseries 05} (2021) 038}, \href{http://arxiv.org/abs/2102.08917}{{\ttfamily arXiv:2102.08917 [hep-th]}}.

\bibitem{hui2022ladder-678}
L.~Hui, A.~Joyce, R.~Penco, L.~Santoni, and A.~R. Solomon, ``{Ladder symmetries of black holes. Implications for love numbers and no-hair theorems},'' \href{http://dx.doi.org/10.1088/1475-7516/2022/01/032}{{\em J. Cosmology Astropart. Phys.} {\bfseries 01} no.~01, (2022) 032}, \href{http://arxiv.org/abs/2105.01069}{{\ttfamily arXiv:2105.01069 [hep-th]}}.

\bibitem{achour2022hidden-8c7}
J.~Ben~Achour, E.~R. Livine, S.~Mukohyama, and J.-P. Uzan, ``{Hidden symmetry of the static response of black holes: applications to Love numbers},'' \href{http://dx.doi.org/10.1007/JHEP07(2022)112}{{\em J. High Energy Phys.} {\bfseries 07} (2022) 112}, \href{http://arxiv.org/abs/2202.12828}{{\ttfamily arXiv:2202.12828 [gr-qc]}}.

\bibitem{charalambous2021hidden-5e0}
P.~Charalambous, S.~Dubovsky, and M.~M. Ivanov, ``{Hidden Symmetry of Vanishing Love Numbers},'' \href{http://dx.doi.org/10.1103/PhysRevLett.127.101101}{{\em Phys. Rev. Lett.} {\bfseries 127} no.~10, (2021) 101101}, \href{http://arxiv.org/abs/2103.01234}{{\ttfamily arXiv:2103.01234 [hep-th]}}.

\bibitem{ivanov2023vanishing-9aa}
M.~M. Ivanov and Z.~Zhou, ``{Vanishing of Black Hole Tidal Love Numbers from Scattering Amplitudes},'' \href{http://dx.doi.org/10.1103/PhysRevLett.130.091403}{{\em Phys. Rev. Lett.} {\bfseries 130} no.~9, (2023) 091403}, \href{http://arxiv.org/abs/2209.14324}{{\ttfamily arXiv:2209.14324 [hep-th]}}.

\bibitem{creci2021tidal-42e}
G.~Creci, T.~Hinderer, and J.~Steinhoff, ``{Tidal response from scattering and the role of analytic continuation},'' \href{http://dx.doi.org/10.1103/PhysRevD.104.124061}{{\em Phys. Rev. D} {\bfseries 104} no.~12, (2021) 124061}, \href{http://arxiv.org/abs/2108.03385}{{\ttfamily arXiv:2108.03385 [gr-qc]}}. [Erratum: Phys.Rev.D 105, 109902 (2022)].

\bibitem{Singha:2025xah}
C.~Singha and S.~Chakraborty, ``{Tidal deformation of black holes in Lovelock gravity},'' \href{http://dx.doi.org/10.1103/8jfh-9rb6}{{\em Phys. Rev. D} {\bfseries 113} no.~2, (2026) 024005}, \href{http://arxiv.org/abs/2508.14944}{{\ttfamily arXiv:2508.14944 [gr-qc]}}.

\bibitem{Rodriguez:2026iot}
M.~J. Rodr{\'\i}guez, L.~Santoni, and A.~R. Solomon, ``{Love numbers of black holes and compact objects},'' \href{http://arxiv.org/abs/2604.08653}{{\ttfamily arXiv:2604.08653 [gr-qc]}}.

\bibitem{Hui:2020xxx}
L.~Hui, A.~Joyce, R.~Penco, L.~Santoni, and A.~R. Solomon, ``{Static response and Love numbers of Schwarzschild black holes},'' \href{http://dx.doi.org/10.1088/1475-7516/2021/04/052}{{\em J. Cosmology Astropart. Phys.} {\bfseries 04} (2021) 052}, \href{http://arxiv.org/abs/2010.00593}{{\ttfamily arXiv:2010.00593 [hep-th]}}.

\bibitem{book:91852226}
R.~K. Nagle, E.~B. Saff, and A.~D. Snider, {\em Fundamentals of Differential Equations and Boundary Value Problems}.
\newblock Pearson Education, Boston, 6~ed., 1993.

\bibitem{Cardoso:2022whc}
V.~Cardoso, K.~Destounis, F.~Duque, R.~Panosso~Macedo, and A.~Maselli, ``{Gravitational Waves from Extreme-Mass-Ratio Systems in Astrophysical Environments},'' \href{http://dx.doi.org/10.1103/PhysRevLett.129.241103}{{\em Phys. Rev. Lett.} {\bfseries 129} no.~24, (2022) 241103}, \href{http://arxiv.org/abs/2210.01133}{{\ttfamily arXiv:2210.01133 [gr-qc]}}.

\bibitem{Figueiredo:2023gas}
E.~Figueiredo, A.~Maselli, and V.~Cardoso, ``{Black holes surrounded by generic dark matter profiles: Appearance and gravitational-wave emission},'' \href{http://dx.doi.org/10.1103/PhysRevD.107.104033}{{\em Phys. Rev. D} {\bfseries 107} no.~10, (2023) 104033}, \href{http://arxiv.org/abs/2303.08183}{{\ttfamily arXiv:2303.08183 [gr-qc]}}.

\bibitem{Dyson:2025dlj}
C.~Dyson, T.~F.~M. Spieksma, R.~Brito, M.~van~de Meent, and S.~Dolan, ``{Environmental Effects in Extreme-Mass-Ratio Inspirals: Perturbations to the Environment in Kerr Spacetimes},'' \href{http://dx.doi.org/10.1103/PhysRevLett.134.211403}{{\em Phys. Rev. Lett.} {\bfseries 134} no.~21, (2025) 211403}, \href{http://arxiv.org/abs/2501.09806}{{\ttfamily arXiv:2501.09806 [gr-qc]}}.

\bibitem{Li:2025ffh}
D.~Li, C.~Weller, P.~Bourg, M.~LaHaye, N.~Yunes, and H.~Yang, ``{Extreme mass-ratio inspiral within an ultralight scalar cloud: Scalar radiation},'' \href{http://dx.doi.org/10.1103/7l9s-g21j}{{\em Phys. Rev. D} {\bfseries 112} no.~8, (2025) 084057}, \href{http://arxiv.org/abs/2507.02045}{{\ttfamily arXiv:2507.02045 [gr-qc]}}.

\bibitem{Brito:2023pyl}
R.~Brito and S.~Shah, ``{Extreme mass-ratio inspirals into black holes surrounded by scalar clouds},'' \href{http://dx.doi.org/10.1103/PhysRevD.108.084019}{{\em Phys. Rev. D} {\bfseries 108} no.~8, (2023) 084019}, \href{http://arxiv.org/abs/2307.16093}{{\ttfamily arXiv:2307.16093 [gr-qc]}}. [Erratum: Phys.Rev.D 110, 109902 (2024)].

\end{thebibliography}\endgroup
\bibliographystyle{./utphys1}

\end{document}